\documentclass[aps, prl, reprint,superscriptaddress]{revtex4-2}

\usepackage{standalone}
\usepackage[bottom]{footmisc}
\usepackage{lipsum}
\usepackage{float}
\usepackage{graphicx}
\usepackage{algorithm}
\usepackage{amssymb, amsmath, amsfonts}
\usepackage[caption=false]{subfig} % <- Preamble
\usepackage{pgfplots} % <- Preamble
\usepackage{xcolor}
\usepackage{multibib}
\usepackage{mathtools}
\usepackage{bm}
\usepackage{amsopn}
\usepackage{xr}
\usepackage{titlesec}
\usepackage{hyperref}
\usepackage[normalem]{ulem} % for crossing out text

\DeclareMathAlphabet{\mathdutchcal}{U}{dutchcal}{m}{n} \SetMathAlphabet{\mathdutchcal}{bold}{U}{dutchcal}{b}{n}
\DeclareMathAlphabet{\mathdutchbcal}{U}{dutchcal}{b}{n}

\makeatletter
\renewcommand*\env@matrix[1][\arraystretch]{%
\edef\arraystretch{#1}%
\hskip -\arraycolsep \let\@ifnextchar\new@ifnextchar \array{*\c@MaxMatrixCols c}}
\newwrite\supptocfile
\immediate\openout\supptocfile=\jobname.suptoc
\makeatletter
\newcommand{\addtosupptoc}[2]{%
  \immediate\write\supptocfile{\string\contentsline {section}{\numberline {#1}#2}{}}
}
\newcommand{\startsupptoc}{%
  \let\oldsection\section
  \renewcommand{\section}[1]{\refstepcounter{section}%
    \addtosupptoc{\thesection}{##1}%
    \oldsection*{\thesection\quad ##1}%
  }
}
\makeatother

\makeatletter
\def\bbordermatrix#1{\begingroup \m@th \@tempdima 4.75\p@ \setbox\z@\vbox{%
    \def\cr{\crcr\noalign{\kern2\p@\global\let\cr\endline}}%
    \ialign{$##$\hfil\kern2\p@\kern\@tempdima&\thinspace\hfil$##$\hfil &&\quad\hfil$##$\hfil\crcr
      \omit\strut\hfil\crcr\noalign{\kern-\baselineskip}%
      #1\crcr\omit\strut\cr}}%
  \setbox\tw@\vbox{\unvcopy\z@\global\setbox\@ne\lastbox}%
  \setbox\tw@\hbox{\unhbox\@ne\unskip\global\setbox\@ne\lastbox}%
  \setbox\tw@\hbox{$\kern\wd\@ne\kern-\@tempdima\left[\kern-\wd\@ne \global\setbox\@ne\vbox{\box\@ne\kern2\p@}%
    \vcenter{\kern-\ht\@ne\unvbox\z@\kern-\baselineskip}\,\right]$}%
  \null\;\vbox{\kern\ht\@ne\box\tw@}\endgroup}
\makeatother

\begin{document}
\raggedbottom
% Use the \preprint command to place your local institutional report number in the upper righthand corner of the title
% page in preprint mode. Multiple \preprint commands are allowed. Use the 'preprintnumbers' class option to override
% journal defaults to display numbers if necessary \preprint{}

%Title of paper
\title{Predicting First-Passage Dynamics in Disordered Systems Exactly: Application to Sparse Networks}
\author{Daniel Marris}
\affiliation{School of Engineering Mathematics and Technology, University of Bristol, Bristol BS8 1TW, United Kingdom}
\author{Chittaranjan Hens}
\affiliation{Center for Computational Natural Science and Bioinformatics, International
Institute of Informational Technology, Hyderabad 500032, India}
\author{Subrata Ghosh}
\affiliation{Center for Computational Natural Science and Bioinformatics, International
Institute of Informational Technology, Hyderabad 500032, India}
\affiliation{Division of Dynamics, Lodz University of Technology, Stefanowskiego 1/15, 90-924 Lodz, Poland}
\author{Luca Giuggioli}
\email{luca.giuggioli@bristol.ac.uk}
\affiliation{School of Engineering Mathematics and Technology, University of Bristol, Bristol BS8 1TW, United Kingdom}

% repeat the \author .. \affiliation  etc. as needed \email, \thanks, \homepage, \altaffiliation all apply to the
% current author. Explanatory text should go in the []'s, actual e-mail address or url should go in the {}'s for \email
% and \homepage. Please use the appropriate macro foreach each type of information

% \affiliation command applies to all authors since the last \affiliation command. The \affiliation command should
% follow the other information \affiliation can be followed by \email, \homepage, \thanks as well.

\date{\today}
\begin{abstract}
Quantifying how spatial disorder affects the movement of a diffusing particle or agent is fundamental to target search studies. When diffusion occurs on a network, that is on a highly disordered environment, we lack the mathematical tools to calculate exactly the temporal characteristics of search processes, instead relying on estimates provided by stochastic simulations. To close this knowledge gap we devise a general methodology to represent analytically the movement and search dynamics of a diffusing random walk on sparse graphs. We show its utility by uncovering the existence of a bi-modality regime in the time-dependence of the first-passage probability to hit a target node in a small-world network. By identifying the network features that give rise to the bi-modal regime, we challenge long-held beliefs on how the statistics of the so-called direct, intermediate, and indirect trajectories influence the shape of the resulting first-passage and first-absorption probabilities and the interpretation of their mean values. Overall these findings show that temporal features in first-passage studies can be utilised to unearth novel transport paradigms in spatially heterogeneous environments.
\end{abstract}
\maketitle

Changes in the movement statistics of an agent as a result of interactions with a disordered environment is a ubiquitous phenomenon. Applications appear both in natural and engineered systems, ranging from the flow of substance through geological or porous substrates \cite{khafagy2022analytical}, the motion of charge carriers in composite battery architectures \cite{ketter2025using}, and the slowing down of colloidal particles in glassy systems \cite{berthier2011theoretical},  to the migration of cells through the extracellular matrix \cite{tsingos2023hybrid} or the dispersal of animals in the landscape \cite{kenkre2021theory}.

The theoretical physics literature devoted to quantify the effects of spatial heterogeneities on the transport of diffusive particles is vast and with a long history \cite{scher1973stochastic}. Some of the recent efforts to analyse these effects have focused on first-passage (FP) times of random walk search in disordered lattices \cite{holehouse2024first, holl2024big, luo2015sample, sarvaharman2023particle}. While early work in this regard has focused on quantifying the mean first-passage time (MFPT) \cite{murthy1989mean,benichou2014first,tejedor2009global}, it is now established that relying exclusively on the MFPT risks overestimating empirical FP times, questioning the use of a single numerical value to characterise search processes \cite{mattos2012first,godec2016first}.  To remedy this a formalism to extract the full FP probability for a disordered 1D lattice has been proposed \cite{holehouse2024first}, but it is only valid for nearest-neighbour jump processes, limiting  its applicability to more general disordered structures. 

To overcome these limitations, we present a general approach that allows to quantify the FP probability in arbitrary disordered lattices. Since networks structures are the most general example of a disordered environment, tools to quantify the FP statistics of a random walker over a network are highly sought after. In network science predicting the time to reach a given node for the first time already serves multiple purposes. From a theoretical perspective it allows to identify influential nodes \cite{gurfinkel2020absorbing} through generalized measures of closeness \cite{white2003algorithms} and betweenness centralities \cite{newman2005measure},  which both consider averages over all random trajectories rather than shortest paths. From the application point of view FP statistics have been used to identify spatial correlations in election results \cite{bassolas2021first} as well as to de-noise \cite{chen2015sparse} and segment images \cite{yu2023techniques}. 

When a random walker is confined to a network, various approaches to study both the FP probability and its mean have been suggested. For arbitrary networks numerical techniques exist, e.g. the renewal analysis~\cite{noh2004random}, the calculation of the adjacency matrix eigenspectrum~\cite{riascossanders2021,giuggioli2022spatio} or iterative procedures~\cite{masuda2017random}. For specific graphs, a recent noteworthy technique is the methodology devised for dense networks \cite{bartolucci2021spectrally} whereby one approximates the MFPT in a computationally efficient way using only the neighbourhood of the target 
location. This approach, however, becomes impractical in sparse networks,  which is expected as spectral methods are known to break down in these cases~\cite{singh2015finding}. %{\color{blue}Chitta/Subrata please check if this sentence is correct.}

One important class of sparse graphs, characterized by their high clustering and short path lengths, is the small-world network (SWN), which are constructed by rewiring regular ring lattices \cite{watts1998collective, amaral2000classes, xu2007small, zarepour2019universal,lachgar2024uncovering}. Numerous works studying the random walk statistics on SWNs have appeared \cite{almaas2003scaling, parris2005traversal, almaas2002characterizing}, some of which play an important role in modelling the propagation of disease and rumours through a population \cite{zanette2002dynamics,almaas2003scaling, wang2022epidemic}. In the context of search statistics, an approximate scaling law for the MFPT as a function of the disorder 
parameter was found through a fitting procedure to numerical data~\cite{pandit2001random}. While such an approach gives good results when the number of nodes is large, it shows considerable deviation from the actual MFPT otherwise \cite{hwang2012first, noh2004random, masuda2017random}.

As many of the challenges to determine with high accuracy the first-passage dynamics relate to the lack of explicit mathematical representation of the diffusive dynamics on networks, we present a general analytical methodology to construct the occupation probability dynamics of a random walk moving randomly on a graph. As our interest is the dynamics on SWNs, we consider an arbitrary $K$-neighbour random walk on the ring lattice, and derive previously unknown expressions for its occupation probability and the MFPT to a single target.

\textit{Dynamics on the ring lattice ---} To study the random walk dynamics on a ring lattice, we start by describing it in an unbounded 1D domain. The Master equation for the occupation probability $P(n,t)$ of a $K = 2k$ neighbour lattice random walk is subject to 
\begin{align}
P(n,t+1)=\frac{1}{K}\sum_{r=1}^k\Big[P(n-r,t)+P(n+r,t)\Big],
\label{eq:kneighbour_lrw}
\end{align}
which states that at each (discrete) time $t$, the walker may jump to site $n$ from $k$ sites to the left or to the right of $n$. We have omitted the probability of staying at site $n$ from the dynamics since we will consider the
Watts-Strogatz SWN network, which typically does not have self-loops. If necessary, one may straightforwardly account for self loops by multiplying 
$q \in (0, 1)$ to the term on the right hand side of Eq. (\ref{eq:kneighbour_lrw}) and by adding the term $(1-q)P(n,t)$. 

To find the dynamics of the occupation probability $Q(n,t)$, where the shift from $P(n,t)$ to $Q(n,t)$ denotes that the random walker is confined to a finite lattice, we follow standard procedures. Namely, we consider a ring of $N$ lattice sites,  and Fourier and $z$-transform  \cite{hughes1996random, montroll1965random} Eq. (\ref{eq:kneighbour_lrw}), before applying the method of images \cite{LucaPRX, hughes1996random} for an initially localised probability $Q(n,0)=\delta_{n,n_0}$. In such a case the so-called propagator solution $Q_{n_0}(n,t)$ of the Master equation, that is Eq. (\ref{eq:kneighbour_lrw}) supplemented by the periodic boundary condition $Q(n,t)=Q(n+N,t)$, is given by the generating function, $\widetilde{Q}_{n_0}(n,z)=\sum_{n=0}^{\infty}Q_{n_0}(n,t)z^t$ \cite{supplemat},
\begin{equation}
    \begin{aligned}
       \widetilde{Q}_{n_0}^{(k)}(n, z) \!=\! \frac{1}{N}\!\sum_{\ell=0}^{N-1}\!\frac{\cos\left(\frac{2\pi\ell(n-n_0)}{N}\right)}{1\!-\!\frac{z}{k} \sin\left(\frac{k \ell \pi }{N}\right)\cos\left(\frac{(k+1)\ell \pi}{N}\right)\csc\left(\frac{\ell \pi}{N}\right)},
    \end{aligned}
    \label{eq: n_z_anyk}
\end{equation}
where we have used the superscript $(k)$ to make explicit the dependence on $k$ (or $K$). The time dependent $Q_{n_0}^{(k)}(n, t)$ is trivially obtained from  (\ref{eq: n_z_anyk}). Note that when $k \in \{1, 2, 3\}$ one may avoid the need to represent the propagator as a finite series by exploiting the connection between certain finite trigonometric series and Chebyshev polynomials \cite{supplemat}. 

With Eq. (\ref{eq: n_z_anyk}), we employ the reactive defect technique to extract the first-absorption probability generating function \cite{giuggioli2022spatio, kenkre2021memory}  as
\begin{align}
\widetilde{A}_{n_0}(n, z) = \rho\, \widetilde{Q}^{(k)}_{n_0}(n, z)/\left[1-\rho + \rho\,\widetilde{Q}^{(k)}_{n}(n, z)\right],
\label{eq:Az}
\end{align}
where $\rho \in (0, 1]$ is the probability of the search being successful, that is the probability  that the walker is absorbed whenever it is located at the target site $n$. From Eq. (\ref{eq:Az}) one deduces 
the FP probability generating function in the limit $\rho = 1$, and the mean first-absorption time  \cite{giuggioli2022spatio} $\mathcal{A}_{n_0\to n}(\rho) = \mathcal{F}_{n_0 \to n} + \frac{1-\rho}{\rho} \mathcal{R}_{n}$, which is trivially related to the MFPT $\mathcal{F}_{n_0 \to n}$ and the mean return time (MRT) to $n$, $\mathcal{R}_{n}$. 

With the propagator in Eq. (\ref{eq: n_z_anyk}), the MFPT and MRT can be extracted algebraically from the probability generating function using renewal relations \cite{redner2001guide, montroll1965random}. For the MFPT we obtain 
\begin{equation}
    \mathcal{F}^{(k)}_{n_0\to n} = k\sum_{\ell=1}^{N-1}\frac{\left[1-\cos\left(\frac{2\pi\ell(n-n_0)}{N}\right)\right]\sin\left(\frac{\ell \pi}{N}\right)}{  k \sin\left(\frac{\ell \pi}{N}\right)-\sin\left(\frac{k \ell \pi }{N}\right)\cos\left(\frac{(k+1)\ell \pi}{N}\right)}.
    \label{eq: MFPT_anyk}
\end{equation}
Note that if self loops are present, i.e. $q\neq 1$, Eq. (\ref{eq: MFPT_anyk}) is modified by a $q^{-1}$ multiplication factor. The MRT is found as
$\mathcal{R}_{n} = N$, regardless of $k$, which is expected since the general MRT on a network may be written as
$\mathcal{R}_{n} = \mathcal{E}/\chi_{n}$, where $\mathcal{E}$ is the number of edges in the network and
$\chi_{n}$ is the degree (co-ordination number) of node $n$ \cite{noh2004random}. 
\begin{figure*}[t]
\includegraphics[width = \textwidth]{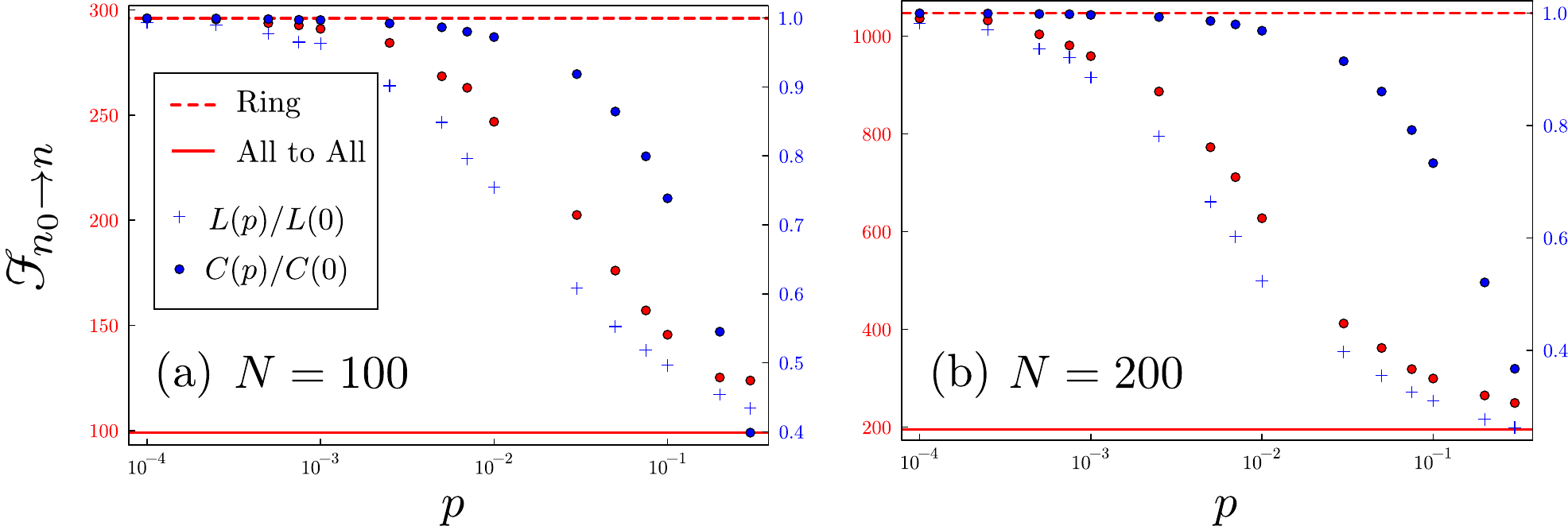}
\caption[A comparison between the MFPT and the graph properties of the characteristic path length and clustering coefficient on small-world
networks.]{A comparison between the MFPT (red axis on the left) and the graph properties of the characteristic path
length, $L(p)$, and clustering coefficient, $C(p)$, (blue axis on the right) on small-world networks with $K = 10$. On the right axis we follow ref. \cite{watts1998collective} and normalise the coefficients by the appropriate coefficient for the ring lattice ($C(0)$ or $L(0)$). For all
MFPT calculations, we take $n_0 = 1$ and $n = N/2$, i.e., the longest path distance on the ring lattice. We also
display the result on the $K=10$ ring graph (dashed line) and the all-to-all graph (solid line). Each point and cross in both
plots represents an ensemble average of 50 realisations of the disorder for each $p$. We determine the
characteristic path length by averaging the shortest path, found via Dijkstra's algorithm \cite{dijkstra1959note}, over
all start and end nodes.}
    \label{fig: mfpt_p_sw}
\end{figure*}
Two well-known limits of Eq. (\ref{eq: MFPT_anyk}) can be recovered. The $k=1$ case reduces to \cite{LucaPRX} $\mathcal{F}^{(1)}_{n_0\to n} = \left(N -|n-n_0|\right)|n-n_0|$. For the opposite limit, the fully connected graph, we take $k =
\frac{N-1}{2}$ ($N \in 2\mathbb{N}+1)$ and we obtain  \cite{bartolucci2021spectrally,pandit2001random} $\mathcal{F}^{\left(\frac{N-1}{2}\right)}_{n_0\to n} = (1-\delta_{n,n_0})(N-1)$ (see \cite{supplemat} for the derivation of the same expression when $N$ is even).

\textit{Dynamics on the network ---} To build the SWN, we follow the Watts-Strogatz prescription \cite{watts1998collective}, whereby a regular ring lattice is
transformed through the introduction of random re-wirings  governed by some probability $p$ \footnote{Both the neighbour to be cut and the node to choose from are drawn from a
uniform distribution over the available nodes.}. As $p$ is increased, the structure of the graph becomes more disordered, approaching a random network as $p\rightarrow 1$.

Each of the rewired links in the SWN represents a spatial \emph{defect} of the inert type (probability preserving) of an otherwise homogenous ring lattice. The number of required defects is dependent on $p$, $N$ and $k$. In particular, the average number of defects for a given 
set of parameters is
\begin{equation}
    M = (K+1)pN,
    \label{eq: num_defects}
\end{equation}
which arises as each SWN has, on average, $pN$ re-wirings \cite{watts1998collective} and each re-wiring requires $K+1$
defects. This ensures that the probability of taking each outgoing edge from site $n$ is maintained as $1/\chi_{n}$, i.e.,
upon cutting or adding a link at $n$, the coordination number of that node has changed, meaning every other edge also
requires its probability to be altered. Hence, the procedure is particularly useful when $K+1<1/p$ i.e., for sparse networks (low $k$) with high clustering
coefficients (small $p$), since the size of the
defect matrices required to find the occupation probability exactly (explained below) is smaller than $N\times N$. 

In the presence of $M$ inert defects the Master equation governing the lattice random walk probability can be expressed by directly altering the transition probabilities via
\cite{sarvaharman2023particle}
\begin{align}
& S(n, t+1) = \sum_{n'}\mathbb{B}_{nn'}S(n', t)  \nonumber \\ &+\sum_{\mathdutchcal{m}=1}^{M}(\delta_{n, u_\mathdutchcal{m}}-\delta_{n, v_\mathdutchcal{m}})\left[\eta_{v_\mathdutchcal{m}, u_\mathdutchcal{m}}S(u_\mathdutchcal{m}, t) - \eta_{u_\mathdutchcal{m}, v_\mathdutchcal{m}}S(v_\mathdutchcal{m}, t)\right],
 \label{eq: defect_ME}
\end{align}
where $\mathbb{B}_{nn'}$
represents the defect-free transition probability from $n'$ to $n$,  $\delta_{a,b}$ is a Kronecker delta, and $\eta_{x, y}$ quantifies the alteration of the probability of jumping from $y$ to $x$. In other words $S(n, t+1) = \sum_{n'}\mathbb{B}_{nn'}S(n', t)$ represents the defect-free Master equation for the lattice walker on the ring lattice, namely Eq. (\ref{eq:kneighbour_lrw}).
\begin{figure*}[t]
    \includegraphics[width = \textwidth]{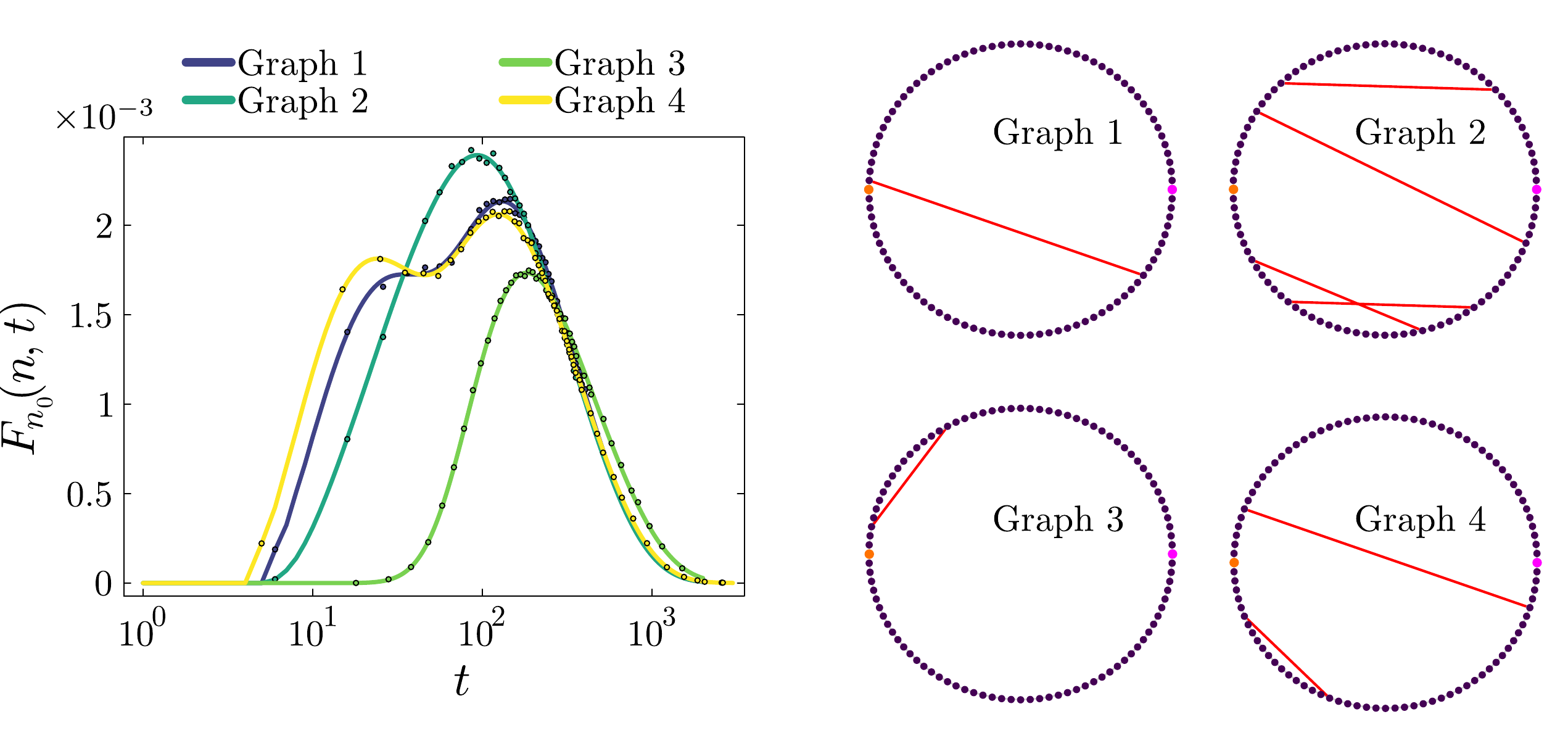}
    \caption[A comparison of the first-passage distributions for four independent realisations of the small-world
    network. Each analytical curve is verified against stochastic simulations.]{A comparison of the FP
    distributions for four independent realisations of the SWN constructed from $K=6$ ring lattice. On the left each analytical curve (solid line) is verified against stochastic simulations (dots), while the schematics of each of the network structure is displayed on the right. For each realisation the network is generated with $p=0.15$ and $N = 100$, and the initial ($n_0=1$) and target ($n=50$) node are shown as slightly larger than the other nodes and depicted, respectively, in orange and pink colour. For ease of visualisation and to give prominence to the different types of
    disorder, we have removed the $K$ neighbour connections from the network schematics and coloured the long range connections red.}
    \label{fig: fp_sw}
\end{figure*}

From Eq. (\ref{eq: defect_ME}), one sees that each re-wiring to build the SWN from the ring lattice requires four pieces of information regarding the inert defects, namely $(u, v, \eta_{v, u}, \eta_{u,v})$, which is conveniently
represented algorithmically as a four-tuple. To extract the set of four-tuples note that each $\mathbb{B}$ is a symmetric circulant matrix  with $A_{ij} = 1/K $ if $0 < |i - j| \bmod N \leq k $, and $ A_{ij} = 0 $ otherwise. Similarly, one may define a transition matrix containing the jump probabilities on the network as $\mathbb{A}$, which is easily extracted via Graph analysis libraries e.g., 
Graphs.jl \cite{fairbanksGraphs2021} or NetworkX \cite{hagberg2008exploring}. Using this formulation we find the modifications, or defect parameters as the non-zero elements of $\mathbb{X} = \mathbb{B}-\mathbb{A}$ where each
four-tuple is $(i, j, \mathbb{X}_{j,i}, \mathbb{X}_{i,j})$. 

With the dynamics on the ring lattice and the set of defects established, the formalism of ref. \cite{sarvaharman2023particle} may now be used to find analytically the occupation probability and the MFPT, 
respectively. To make conspicuous the change from the ring lattice to the network we use $\mathdutchcal{F}_{n_0\rightarrow n}$ instead of $\mathcal{F}_{n_0\rightarrow n}$ to denote the MFPT on the SWN. The propagator generating function of Eq. (\ref{eq: defect_ME}) and $\mathdutchcal{F}_{n_0\rightarrow n}$ are expressed analytically in terms of determinants of $M \times M$ matrices, whose exact analytic dependence are given explicitly in ref. \cite{supplemat}, and comprise solely expressions pertaining to the ring lattice, respectively Eqs. (\ref{eq: n_z_anyk}) and (\ref{eq: MFPT_anyk}), and the various $\eta$ parameters.

In a SWN, with rewired links that massively cut the typical separation between two vertices, we expect that a portion of the lattice walk trajectories no longer traverse the entire outer ring to reach a given target \cite{pandit2001random}. This effect is visible in Fig. \ref{fig: mfpt_p_sw}, where we plot the MFPT as a function of the re-wiring parameter $p$. As a quantitative comparison we show explicitly that a random walk average search time lies between two limits, the MFPT on the ring lattice
and the one on a fully connected graph. Those two extreme values are bridged  through a rapid decline occurring at the same $p$-threshold for which the characteristic path length drops suddenly, i.e. at the onset of the small-world property.  
\begin{figure*}[t]
    \includegraphics[width = \textwidth]{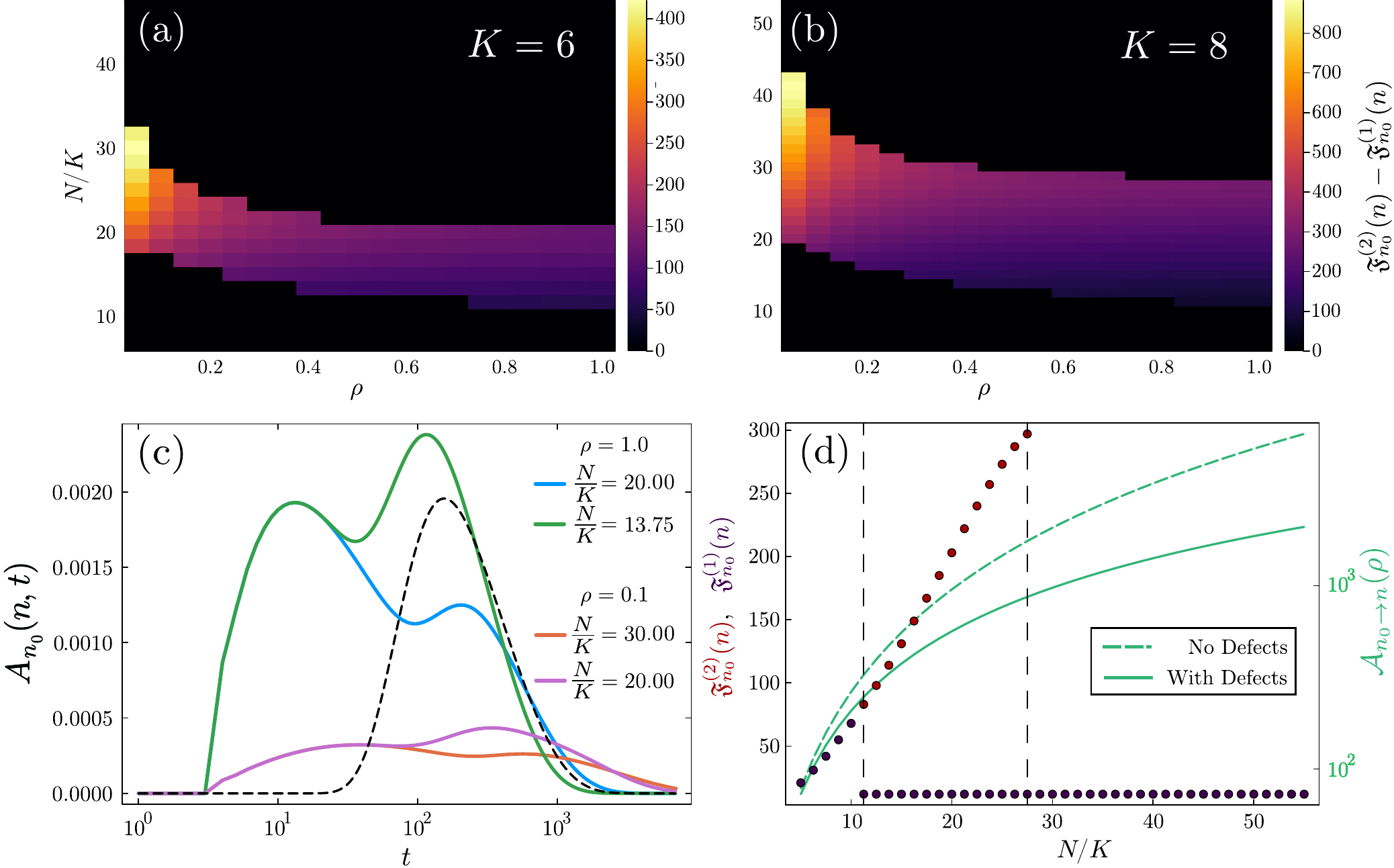}
    \caption[A quantification of when the first-absorption distribution is bi-modal in a disordered ring network as a
    function of the absorption parameter and network size.]{ Bi-modality in the first-absorption distribution from $n_0=1$ to a target at $n=N/2$ with $N$ even on an artificially generated small world network (see text), obtained by replacing $\widetilde{Q}^{(k)}_{n_0}(n,z)$ with the solution $\widetilde{S}_{n_0}(n,z)$ of (\ref{eq: defect_ME})  in Eq. (\ref{eq:Az}) and inverting the resulting generating function to time.  In panels (a) and (b) non-black coloured cells indicate the presence of two modes with the colour intensity representing the amount of time between the locations of the two modes ($\mathfrak{F}^{(i)}_{n_0}(n)$
    denotes the time of the $i^{\text{th}}$ mode). Panel (c) shows examples of the entire distribution for different network parameters in the bi-modal regime, with the defect-free case (no re-wiring) when $\rho=1$ as a dashed curve. Panel (d) compares the dependence of the MFPT, that is $\mathcal{A}_{n_0\rightarrow n}(\rho=1)$, and the first and second mode, which are both present in between the dashed vertical lines, as a function of the network size $N$ ($K=8$). Across all plots we identify the presence of a mode
as a time $t$ when $F_{n_0}(n, t-1) < F_{n_0}(n, t)$ and $F_{n_0}(n, t) > F_{n_0}(n, t+1)$. To avoid potential misclassifications 
we demand that a peak satisfies the inequality $F_{n_0}(n, t) > 10^{-7}$ and, if a second mode exists, it must be
at least $1\%$ of the height of the highest one.}
    \label{fig: network_bi}
\end{figure*}

As the MFPT hides much of the nuances of the search behaviour, in Fig. \ref{fig: fp_sw} we explore the FP
distribution for four realisations of the SWN with the same number of nodes and edges, the same rewiring parameter, and the same initial and target location. The variation in transport characteristics between the four SWNs manifests itself in the qualitative differences of the FP distributions.
%which arise when two distinct timescales appear, the faster one corresponding to a walker exploiting convenient rewired links to reach the target very quickly.
This is particularly evident in Graph 4 where the FP distribution is bi-modal. In that case those walkers that traverse the shortcut early on and then reach the target characterise the \emph{direct} trajectories \cite{godec2016first} and give rise to the first mode of the FP probability. While the tail of the FP distribution is controlled by the so-called \emph{indirect} trajectories with long excursion across the entire domain, there exists a temporal regime with trajectories, referred to as \emph{intermediate}, that make initial excursions away from the target, before proceeding directly to it \cite{godec2016universal}. We posit that a pronounced timescale separation between direct and intermediate trajectories and the presence of strong spatial heterogeneities that enhance the latter is what leads to a regime of bi-modality in the FP probability.

To confirm the above assertion, we artificially create a network with a 
structure similar to Graph 4 of Fig. \ref{fig: fp_sw}. We do that by taking a ring network with a given $K$ and introducing a `helpful' connection by placing one short-cut between nodes $n+1$ and $n_0 + 5$, where $n$ is
the target node, thus having $M=K+1$ defects in Eq. (\ref{eq: defect_ME}). In this way, by engineering a connection that links the target and initial node with a small number of steps, we control the direct trajectories and thus the timescale of the first mode.
Keeping this local network structure, we study how the
number of nodes, $N$, connectivity $K$, and the absorption parameter at the target, $\rho$, affect the intermediate trajectories and, in turn, the appearance of a second mode in the FP probability, as displayed in Fig.  \ref{fig: network_bi}(c) for certain parameter choice (see also Fig. S2 in \cite{supplemat}).

In Figs. \ref{fig: network_bi}(a) and \ref{fig: network_bi}(b) we identify with coloured cells the regime of bi-modality when a walker is absorbed at $n$ starting from $n_0$. The black areas in the panels correspond to when the FP probability is unimodal, where,  the FP mode at the top being always smaller, for the same value of $\rho$, than the one at the bottom. The area in between corresponds to when both modes are present and points to two features for their simultaneous appearance. On one hand the ratio $N/K$ needs to be large enough (above the bottom black region) such that most direct FP trajectories, which give rise to the first mode, $\mathfrak{F}^{(1)}_{n_0}(n)$, are absorbed before those that traverse the outer ring, the latter ones giving rise to the second mode, $\mathfrak{F}^{(2)}_{n_0}(n)$. On the other hand $N/K$ cannot be too big (below the top black region) to avoid large variation in search times of the outer intermediate trajectories that would make the second mode disappear. See Fig. S2 in ref. \cite{supplemat} for a visual comparison of these distribution types.

As an imperfect absorption ($\rho<1$) has the effect of extending the length of all FP trajectories, the two FP modes get reduced in magnitude and become broader the smaller the value of $\rho$. For a given ratio $N/K$, by taking progressively smaller values of  $\rho$, the trajectories taking the shortcut wander for increasingly longer time before being absorbed at the target. As the absorption timescale for such trajectories acquire large variability and become comparable to some of the movement paths that go around the outer ring, for a set $N/K$, below a threshold value of $\rho$ there is no second mode. In such scenario, without changing $\rho$, bi-modality can be restored by increasing the size of the network, which allows to bring back the temporal separation between direct and intermediate trajectories. In fact, as larger $K$ values increase variability in FP times, one may still observe two modes but needs to have a network with a larger number of nodes, which in turn makes absorption times for the outer trajectories take longer. These are the reasons why, in comparing panels (a) and (b), we observe an upward shift, a broadening of the bi-modality region and an increase in timescale separation between $\mathfrak{F}^{(2)}_{n_0}(n)$ and $\mathfrak{F}^{(1)}_{n_0}(n)$.

Past studies of heterogeneity-controlled kinetics, which uncovered the existence of a third timescale associated with intermediate trajectories, have concluded that the MFPT, while being dominated by indirect trajectories, is completely specified by the statistics of direct trajectories \cite{godec2016universal}. To bring evidence that this statement is not universally valid, in Fig.\ref{fig: network_bi}(d) we show that in the bi-modal region the MFPT does increase as the network size increases, but without any variation in $\mathfrak{F}^{(1)}_{n_0}(n)$, that is leaving the statistics of direct trajectories unchanged (for $\rho\not= 1$ see \cite{supplemat}). For sufficiently small $N/K$ values we also observe that $\mathfrak{F}^{(2)}_{n_0}(n)$ becomes the largest characteristic timescales in the system, even larger than the MFPT. Given the non-homogeneous nature of our small world network, as compared to the radially symmetric arrangement studied in ref. \cite{godec2016universal}, we deduce that in strongly heterogeneous environments, only two typical FP timescales may appear, the early one associated with direct trajectories, and the later one associated with intermediate trajectories, and that the MFPT is not only independent of the statistics of direct trajectories, but loses its meaning as representing the average time for the indirect trajectories to be absorbed at the target.

\textit{Discussion ---} In this Letter we provide a general framework to study analytically the first-passage probability on network structures devising a procedure to exploit the inert defect technique \cite{sarvaharman2023particle}. For our application to SWNs we derive previously unknown expressions for the dynamics of the probability of the $k$ nearest-neighbour random walk. In doing so, we show that the spatio-temporal search dynamics of a random walker on a SWN is highly sensitive to the network structure. We uncover, for the first time in symmetric and diffusive systems, bi-modal first-absorption distributions, which are driven by the ratio of the network size and the number of nearest neighbours in the ring lattice. We also provide evidence that the mode at short times is dependent on the local network structure around the initial condition and target location, that the later mode is dependent on the global network structure, and that in the bi-modal regime the MFPT is not affected by the statistics of the direct trajectories and cannot be interpreted as the characteristic timescale of indirect trajectories.

The strength of our approach lies in its general applicability. As input it requires mathematical knowledge of the dynamics of the occupation probability in a topology that is not too dissimilar from the specific problem at hand. Such scenario occurs in any graph that displays structural similarities to a strongly regular graph \cite{brouwer2012strongly} or to some other specific graph over which the dynamics of the random walk occupation probability is known analytically. In these situations the entire time-dependent probability of the walk as well of the first-passage probability can be extracted without recurring to stochastic simulations. While the methodology becomes computationally onerous if the topology is very different from that of the associated defect-free case, that is when the number of defects become large, it is however valid irrespective of the strength of the defect. It is thus an effective approach when a walker interacts strongly with the disorder in the environment and ideally suited to link temporal features to spatial heterogeneity in FP processes.

\textit{Acknowledgments}—DM acknowledges support from the Engineering and Physical Research Council (EPSRC) DTP studentship, grant number EP/T517872/1, and LG from the National Environment Research
Council (NERC) grant number NE/W00545X/1. CH acknowledges the support of a Bristol Next Generation Visiting Researcher fellowship from the University of Bristol to visit and conduct part of the present work.

\bibliography{references}

\begin{thebibliography}{10}
\bibitem{supplemat}
See {S}upplemental {M}aterial at [url will be inserted by publisher] for
  further details on the derivations.
\end{thebibliography}
\bibliographystyle{unsrt}

\pagebreak
\onecolumngrid

\begin{center}
\textbf{\large Supplemental Materials: Predicting First-Passage Dynamics in Disordered Systems Exactly: Application to Sparse Networks}

\end{center}
%%%%%%%%%% Merge with supplemental materials %%%%%%%%%% %%%%%%%%% Prefix a "S" to all equations, figures, tables and
%reset the counter %%%%%%%%%%
\setcounter{equation}{0}
\setcounter{figure}{0}
\setcounter{table}{0}
\setcounter{page}{1}
\makeatletter
\renewcommand{\theequation}{S\arabic{equation}} \renewcommand{\thefigure}{S\arabic{figure}}
% Section/figure/table/equation numbering
\renewcommand{\thesection}{\Roman{section}}
\renewcommand{\thesubsection}{\thesection.\Alph{subsection}}
\renewcommand{\thesubsubsection}{\thesubsection.\arabic{subsubsection}}
\renewcommand{\thefigure}{S\arabic{figure}}
\renewcommand{\thetable}{S\arabic{table}}
\renewcommand{\theequation}{S\arabic{equation}}

% Show section numbers in TOC and bookmarks
\setcounter{secnumdepth}{3}
\setcounter{tocdepth}{3}
%%%%%%%%%% Prefix a "S" to all equations, figures, tables and reset the counter %%%%%%%%%%
\tableofcontents
\section{Occupation probability of the $K$-neighbour walk}
The general dynamics on the $K$-neighbour ring lattice may be found by considering first the unbounded dynamics along lattice site $n$ given by via the Master equation
\begin{align}
P(n,t+1)=\frac{q}{K}\sum_{r=1}^k\Big[P(n-r,t)+Q(n+r,t)\Big] + (1-q)P(n, t),
\label{eq_sup:kneighbour_lrw}
\end{align}
where Eq. (1) of the main text is obtained by taking $q=1$. Equation (\ref{eq_sup:kneighbour_lrw}) may be solved by performing the Fourier transform, $\widehat{f}(\xi) = \sum_{n=-\infty}^{\infty}f(n)\text{e}^{i \xi n}$,
\begin{equation}
    \widehat{P}^{(k)}(\xi, t+1) = \left[1-q+\frac{q}{K}\sum_{r=1}^{k}\cos(r\xi)\right]\widehat{P}^{(k)}(\xi, t) ,
    \label{eq_sup: fourier_anyk}
\end{equation}
where we use the superscript $k$ to indicate the number 
of left and right neighbours ($K=2k$). After $z$-transforming the time variable, $\widetilde{f}(z) = \sum_{t=0}^{\infty}z^t f(t)$, and by taking the inverse Fourier transform, we obtain the Lattice Green's Function, or propagator,
\begin{equation}
    \widetilde{P}^{(k)}_{n_0}(n, z) = \frac{1}{2\pi}\int_{-\pi}^{\pi}\frac{e^{-i \xi (n-n_0)}}{1-z\left[1-q + \frac{q}{k}\sum_{r=1}^{k}\cos(r \xi)\right]}\text{d}\xi,
    \label{eq_sup: fourier_z_anyk}
\end{equation}
where $n_0$ is the walker's initial site ($\widehat{P}^{(k)}(\xi,0)=e^{i\xi n_0}$).

To simplify Eq. (\ref{eq_sup: fourier_z_anyk}) we use the identity $\sum_{r=1}^{k}\cos(r \xi) = \sin\left(\frac{k
\xi}{2}\right)\cos\left(\frac{(k+1)\xi}{2}\right)\csc\left(\frac{\xi}{2}\right)$
to obtain 
\begin{equation}
    \widetilde{P}^{(k)}_{n_0}(n, z) = \frac{1}{2\pi}\int_{-\pi}^{\pi}\frac{e^{-i \xi (n-n_0)}}{1-z\left[1-q + \frac{q}{k}\sin\left(\frac{k
\xi}{2}\right)\cos\left(\frac{(k+1)\xi}{2}\right)\csc\left(\frac{\xi}{2}\right)\right]}\text{d}\xi.
    \label{eq_sup: fourier_z_anyk1}
\end{equation}
To bound the dynamics to a ring lattice we apply the method of images $\widetilde{Q}_{n_0}^{(k)}(n, z) = \sum_{\ell =\infty}^{\infty}\widetilde{P}^{(k)}_{n_0+ \ell N}(n, z)$ \cite{hughes1996random, montroll1965random, LucaPRX}, where $N$ is the number of sites in the finite lattice, leading to 
\begin{equation}
    \widetilde{Q}^{(k)}_{n_0}(n, z) = \frac{1}{2\pi}\int_{-\pi}^{\pi}\frac{\sum_{\ell = -\infty}^{\infty}e^{-i \xi (n-n_0-\ell N)}}{1-z\left[1-q + \frac{q}{k}\sin\left(\frac{k
\xi}{2}\right)\cos\left(\frac{(k+1)\xi}{2}\right)\csc\left(\frac{\xi}{2}\right)\right]}\text{d}\xi.
    \label{eq_sup: fourier_z_anyk2}
\end{equation}
To solve the integral in Eq. (\ref{eq_sup: fourier_z_anyk2}) we employ the identity \cite{hughes1996random, lighthill1958introduction}
\begin{equation}
    \sum_{\ell = -\infty}^{\infty}e^{-i \ell N\xi } = \frac{2\pi}{N}  \sum_{\ell = -\infty}^{\infty} \delta\left(\xi - \frac{2 \pi \ell}{N}\right),
    \label{eq_sup: dirac}
\end{equation}
to perform the integration and find (see ref. \cite{hughes1996random} for a detailed discussion of integrals of this type)
\begin{equation}
    \widetilde{Q}_{n_0}^{(k)}(n, z) = \frac{1}{N(1-z)} + \frac{1}{N}\sum_{\ell=1}^{N-1}\frac{\cos\left(\frac{2\pi\ell(n-n_0)}{N}\right)}{1-\frac{z}{k} \left[1-q + \frac{q}{k}\sin\left(\frac{k
\ell \pi}{N}\right)\cos\left(\frac{(k+1)\ell \pi}{N}\right)\csc\left(\frac{\ell\pi}{N}\right)\right]}.
\label{eq_sup: prop_nk}
\end{equation}
Finally, noting that $\lim_{\ell \to 0 }\sin\left(\frac{k \ell \pi }{N}\right)\cos\left(\frac{(k+1)\ell \pi}{N}\right)\csc\left(\frac{\ell \pi}{N}\right) = 1$ and also taking $q=1$, gives Eq. (2) of the main text.

If required, the time dependent dynamics are easily extracted from Eq. (\ref{eq_sup: prop_nk}) via the inverse $z$-transform, $f(t) = (2\pi i)\oint\widetilde{f}(z)z^{-t-1}\text{d}z$ with $|z|<1$ and the integration contour being counterclockwise. Performing this integral leads to 
\begin{equation}
    Q_{n_0}^{(k)}(n, t) = \frac{1}{N}\sum_{\ell=0}^{N-1}\cos\left(\frac{2\pi\ell(n-n_0)}{N}\right)\left[1-q + \frac{q}{k}\sin\left(\frac{k
\ell\pi}{N}\right)\cos\left(\frac{(k+1)\ell\pi}{N}\right)\csc\left(\frac{\ell\pi}{N}\right)\right]^t.
\label{eq_sup: prop_nk_t}
\end{equation}
The steady state, found from Eq. (\ref{eq_sup: prop_nk}) via $Q^{(k)}(n, t\to \infty) = \lim_{z\to 1}(1-z)Q^{(k)}(n, z)$, is as expected $Q^{(k)}(n, t\to \infty) = 1/N$ for all $n$ and is independent of $k$.
\subsection{Mean squared displacement for the unbounded $K$-neighbour walk}
To understand the effects on the transport process the higher the value of $k$, we derive the time dependence of the mean square displacement (MSD), $\Delta n(t)=\langle n^2 \rangle(t)-\langle n \rangle^2 (t)$, in unbounded space. Solving Eq. (\ref{eq_sup: fourier_anyk}) we find 
\begin{equation}
    \widehat{P}^{(k)}_{n_0}(\xi, t) = \left[1-q+\frac{q}{k} \sin\left(\frac{k \xi}{2}\right)\cos\left(\frac{(k+1)\xi}{2}\right)\csc\left(\frac{\xi}{2}\right)\right]^te^{i\xi n_0},
    \label{eq_sup: k-t}
\end{equation}
where we have taken $P^{(k)}(n,0)=\delta_{n,n_0}$ with $\delta_{a,b}$ the Kronecker delta. Using the fact that $\langle n^2 \rangle(t) =\left.- \frac{\partial ^2 \widehat{P}_{n_0}^{(k)}(\xi,
t)}{\partial \xi^2}\right|_{\xi = 0}$ and $\langle n \rangle (t)=\left.- i\frac{\partial  \widehat{P}_{n_0}^{(k)}(\xi,
t)}{\partial \xi}\right|_{\xi = 0}$, the MSD in the unbounded lattice,
\begin{equation}
    \Delta n(t)  = \frac{q}{6} (1 + k) (1 + K)\,t,
\end{equation}
shows that more long range connections speeds up the diffusive spread with a coefficient proportional to $1+3k+2k^2$. 
\subsection{Representing the propagator on a finite ring via Chebyshev polynomials}
As mentioned in the main text, for $k \in \{1, 2, 3\}$ one may bypass the need to perform the  finite sum in Eq. (\ref{eq_sup: prop_nk}). Below we show the derivations for each of the case.

\subsubsection{The nearest-neighbour ($k=1)$ walk}
This case has been derived in ref. \cite{LucaPRX}, and we thus simply state the result here:
\begin{equation}
    \widetilde{Q}^{(1)}_{n_0}(n, z) = \frac{\cosh\Big[\Big(N-|n-n_0|\Big)\gamma\Big] + \cosh\Big[|n-n_0|\gamma\Big]}{zq \sinh\left(\gamma\right)\sinh\left(N\gamma\right)}
\end{equation}
where $\cosh\left(\gamma\right) = 1 + \frac{1}{q}\left[\frac{1}{z}-1\right]$ and $\sinh\left(\gamma\right) =\frac{ \sqrt{(1-z)\big[1-(1-2q)z\big]}}{qz}$.

\subsubsection{The next-nearest-neighbour ($k=2)$ walk}
When $k=2$ we rewrite the unbounded propagator in Eq. \ref{eq_sup: fourier_z_anyk} as
\begin{align}
\widetilde{P}^{(2)}_{n_0}(n,z)&=\frac{1}{2\pi}\int_{-\pi}^{\pi}\frac{e^{-i\xi(n-n_0)}}{1-z\left[1-\frac{3q}{2}+q\cos^2(\xi)+\frac{q}{2}\cos(\xi)\right]}\text{d}\xi, \nonumber \\
        &=\frac{1}{2\pi}\int_{-\pi}^{\pi}\frac{e^{-i\xi(n-n_0)}}{2zq\sqrt{\frac{25}{16}+\frac{1}{q}\left[\frac{1}{z}-1\right]}}\left\{ \frac{1}{\beta^+_z}\frac{1}{1+\frac{\cos(\xi)}{\beta^+_z}}+\frac{1}{\beta^-_z}\frac{1}{1-\frac{\cos(\xi)}{\beta^-_z}}\right\}\text{d}\xi,
\label{eq_sup:sol_nnn_Four}
\end{align}
where $|\beta^{\pm}_z|>1$ defined by
\begin{align}
\beta^{\pm}_z=\sqrt{\frac{25}{16}+\frac{1}{q}\left[\frac{1}{z}-1\right]}\pm \frac{1}{4},
\label{eq:betapm}
\end{align}
and where the second line in (\ref{eq_sup:sol_nnn_Four}) has been obtained by finding the zeros of the quadratic polynomial in $\cos(\xi)$ and making a partial fraction expansion. This expansion enables us to compute the integral to obtain
\begin{align}
\widetilde{P}^{(2)}_{n_0}(n,z)=\frac{1}{2zq\left(\beta_{z}^{+}-\frac{1}{4}\right)}\left\{\frac{1}{\sqrt{(\beta_z^+)^2-1}}\frac{(-1)^{n-n_0}}{\left(\sqrt{(\beta^+_z)^2-1}+\beta_z^+\right)^{|n-n_0|}}+\frac{1}{\sqrt{(\beta_z^-)^2-1}}\frac{1}{\left(\sqrt{(\beta^-_z)^2-1}+\beta_z^-\right)^{|n-n_0|}}\right\}.
\label{eq_sup:sol_nnn_space}
\end{align}
One may verify normalisation via $\sum_{n=-\infty}^{+\infty}\widetilde{Q}^{(2)}_{n_0}(n,z)=\frac{1}{1-z}$ and recover the appropriate initial condition, i.e. by taking the $z\rightarrow 0$ limit that $\widetilde{Q}^{(2)}_{n_0}(n,0)=\delta_{n.n_0}$. One may also identify how the probability decays to zero al long times by looking at the $z\rightarrow 1$ limit, which gives $\widetilde{Q}^{(2)}_{n_0}\simeq\frac{1}{\sqrt{5q}}\frac{1}{\sqrt{1-z}}$, and corresponds to $Q^{(2)}_{n_0}(n,t)\simeq\frac{1}{\sqrt{5\pi q\, t}}$ for $t\rightarrow \infty$. %\sout{Thus one can see that the expected long time dependence is satisfied through the final value theorem: $(1-z)\widetilde{Q}(n,z)\rightarrow 0$.}

To bound the domain we again employ the method of images shown in Eq. (\ref{eq_sup: fourier_z_anyk2}), using Eq. (\ref{eq_sup: dirac}), to deduce
%\begin{equation}
%    \widetilde{Q}^{(2)}_{n_0}(n,z)=\frac{1}{2\pi}\int_{-\pi}^{\pi}\frac{\text{e}^{-i\xi(n-n_0)}}{2zq\sqrt{\frac{25}{16}+\frac{1}{q}\left[\frac{1}{z}-1\right]}}\left\{ \frac{1}{\beta^+_z}\frac{1}{1+\frac{\cos(\xi)}{\beta^+_z}}+\frac{1}{\beta^-_z}\frac{1}{1-\frac{\cos(\xi)}{\beta^-_z}}\right\}\text{d}\xi,
%\label{eq_sup:sol_nnn_Four1}
%\end{equation}
\begin{align}
&\widetilde{Q}^{(2)}_{n_0}(n,z)=\frac{1}{2zq\sqrt{\frac{25}{16}+\frac{1}{q}\left[\frac{1}{z}-1\right]}N}\left\{\sum_{\ell=0}^{N-1}\frac{\cos\left(\frac{2\pi\ell(n-n_0)}{N}\right)}{\beta_z^--\cos\left(\frac{2\pi\ell}{N}\right)}+\sum_{\ell=0}^{N-1}\frac{\cos\left(\frac{2\pi\ell(n-n_0)}{N}\right)}{\beta_z^++\cos\left(\frac{2\pi\ell}{N}\right)}\right\} \nonumber \\
&=\frac{1}{N(1-z)}+\frac{1}{2zq\sqrt{\frac{25}{16}+\frac{1}{q}\left[\frac{1}{z}-1\right]}N}\left\{\sum_{\ell=1}^{N-1}\frac{\cos\left(\frac{2\pi\ell(n-n_0)}{N}\right)}{\beta_z^--\cos\left(\frac{2\pi\ell}{N}\right)}-\sum_{\ell=1}^{N-1}\frac{\cos\left(\frac{2\pi\ell(n-n_0)}{N}\right)}{-\beta_z^+-\cos\left(\frac{2\pi\ell}{N}\right)}\right\} \nonumber \\
%&=\frac{1}{2zq\sqrt{\frac{25}{16}+\frac{1}{q}\left[\frac{1}{z}-1\right]}}\left\{  \frac{\cosh\Big[(N-|n-n_0|)\zeta^-\Big]+\cosh\Big[|n-n_0|\zeta^-\Big]}{\sqrt{(\beta_z^-)^2-1}\sinh[N\zeta^-]}\right.\nonumber \\
%&\left.\frac{(-1)^{|n-n_0|}\left\{\cosh\Big[(N-|n-n_0|)\zeta^+\Big]+(-1)^N\cosh\Big[|n-n_0|\zeta^+\Big]\right\}}{\sqrt{(\beta_z^+)^2-1}\sinh[N\zeta^+]}\right\} \nonumber \\
&=\frac{1}{2zq\sqrt{\frac{25}{16}+\frac{1}{q}\left[\frac{1}{z}-1\right]}}\left\{\frac{T_{N-|n-n_0|}(\beta^{-}_z)+T_{|n-n_0|}(\beta^{-}_z)}{\left[(\beta^-_z)^2-1\right]U_{N-1}(\beta^-_z)} +(-1)^{n-n_0}\frac{T_{N-|n-n_0|}(\beta^{+}_z)+(-1)^NT_{|n-n_0|}(\beta^{+}_z)}{\left[(\beta^+_z)^2-1\right]U_{N-1}(\beta^+_z)}\right\},
\label{eq_sup:sol_nnn_space_2}
\end{align}
where in the last equation trigonometric identities from ref. \cite{LucaPRX} have been used with  $T_r(x)$ and $U_r(x)$ indicating Chebyshev polynomials, respectively, of the first and second kind, of degree $r$ and argument $x$.
%where I have used Chebyshev properties and the finite series identities derived in the PRX paper to get to the last line with $$\zeta^{\pm}=\mbox{arccosh}(\beta_z^{\pm})$$
% Note that this provides an interesting paired Chebyshev identity between the $k=2$ case in Eq. (\ref{eq_sup: prop_nk}) and Eq. (\ref{eq_sup:sol_nnn_space_2}).
% Note that the corresponding expression for $P_{n_0}(n,t)$, that is the inverse generating function of Eq. (\ref{eq_sup:sol_nnn_space_2}), is simply given by
% \begin{align}
% P_{n_0}(n,t)=\frac{1}{N}\sum_{\ell=0}^{N-1}\cos\left[\frac{2\pi\ell(n-n_0)}{N}\right]\left[1-\frac{3q}{2}+q\cos^2\left(\frac{2\pi\ell}{N}\right)+\frac{q}{2}\cos\left(\frac{2\pi\ell}{N}\right)\right]^t,
% \label{eq_sup:k2_timedepP}
% \end{align}
% which provides an interesting paired Chebyshev identity (confirmed numerically) between the last expression of Eq. (\ref{eq_sup:sol_nnn_space_2}) and (\ref{eq_sup:k2_timedepP}).

\subsubsection{The next-next-nearest-neighbour ($k=3$) walk}
When $k = 3$ the generating function of the Fourier unbounded propagator in Eq. (\ref{eq_sup: k-t}) can be written as 
\begin{align}
\widetilde{\widehat{P}}^{(3)}_{n_0}(\xi,z)&=\frac{e^{i\xi n_0}}{1-z\left\{1-q+\frac{q}{3}\Big[\cos(3k)+\cos(2k)+\cos(\xi)\Big]\right\}} \nonumber \\
&= \frac{e^{i\xi n_0}}{1-z\left\{1-\frac{4q}{3}+\frac{4q}{3}\left[\cos^3(\xi)+\frac{1}{2}\cos^2(\xi)-\frac{1}{2}\cos(\xi)\right]\right\}} \nonumber \\
&=\frac{3}{4zq}\frac{e^{i\xi n_0}}{1+\frac{3}{4q}\left[\frac{1}{z}-1\right]-\Big[\cos^3(\xi)+\frac{1}{2}\cos^2(\xi)-\frac{1}{2}\cos(\xi)\Big]} \nonumber \\
&=\frac{3}{4zq}\left\{\frac{A_0}{\lambda_0-\cos(\zeta)}+\frac{B_+}{\lambda_+-\cos(\zeta)}+\frac{B_-}{\lambda_--\cos(\zeta)} \right\}e^{i\xi n_0},
\label{eq:unb_Four_k3}
\end{align}
where the $\lambda$'s are the roots of the cubic equation in $\cos(\xi)$, namely
\begin{align}
\lambda_0&=-\frac{1}{6}-\frac{7\zeta^{-1}+\zeta}{6}, \nonumber \\
\lambda_{\pm}&=-\frac{1}{6}+\frac{7\zeta^{-1}+\zeta}{12}\pm i\sqrt{3}\frac{7\zeta^{-1}-\zeta}{12},
\label{eq:def_lam}
\end{align}
with
\begin{align}
\zeta=\left\{10-108\left[1+\frac{3}{4q}\left(\frac{1}{z}-1\right)\right]+\frac{9}{qz}\sqrt{3^4(1-z)^2+\frac{7^3}{3} q^2z^2+14^2qz(1-z)}\right\}^{1/3}.
\label{eq:csi}
\end{align}

%\begin{align}
%\lambda_0&=-\frac{1}{6}+\frac{\sqrt{7}}{3} \cosh\left(\phi\right), \\ \nonumber 
%\lambda_{\pm}&=-\frac{1}{6}-\frac{\sqrt{7}}{6}  \left[\cosh\left(\phi\right)\mp i\sqrt{3} \sinh\left(\phi\right)\right],\\ \nonumber 
%\phi&=\frac{1}{3}\mbox{arccosh}\left(\frac{108\,\alpha-10}{7^{\frac{3}{2}}}\right),\qquad \alpha=1+\frac{3}{4q}\left[\frac{1}{z}-1\right].
%\label{eq_sup:eigen_3}
%\end{align}
%\begin{align}
%\lambda_0&=-\frac{1}{3}\left[\frac{1}{2}+\beta_z+\frac{7}{4\beta_z}\right], \nonumber \\
%\lambda_{\pm}&=-\frac{1}{3}\left[\frac{1}{2}-\frac{1}{2}\left(\beta_z+\frac{7}{4\beta_z}\right)\pm i\frac{\sqrt{3}}{2}\left(\beta_z-\frac{7}{4\beta_z} \right) \right],\nonumber  \\
%\beta_z&=\left\{\frac{1}{2}\left[\frac{5}{4}-3^3\alpha+\sqrt{\left(\frac{5}{4}-3^3\alpha\right)^2-\frac{7^3}{4^2}}\right]\right\}^{\frac{1}{3}}, \nonumber \\
%\alpha=&1+\frac{3}{4q}\left[\frac{1}{z}-1\right].
%\end{align}
Given that $\lambda_0+\lambda_++\lambda_-=\lambda_+\lambda_-+\lambda_0(\lambda_++\lambda_-)=-\frac{1}{2}$ and $\lambda_0\lambda_+\lambda_-=1+\frac{3}{4q}\left(\frac{1}{z}-1\right)$ one can show that %=\left[10+7^{3/2}\cosh(3\phi)\right]/108=\alpha$ and
\begin{eqnarray}
A_0&=&\frac{1}{(\lambda_0-\lambda_+)(\lambda_0-\lambda_-)}=\frac{12\zeta^2}{\zeta^4+7\zeta^2+7^2}, \nonumber \\ %\frac{12}{7\left[1+2\cosh\left(2\phi\right)\right]},
B_{\pm}&=&\frac{1}{(\lambda_{\pm}-\lambda_0)(\lambda_{\pm}-\lambda_{\mp})}=\frac{6\zeta^2}{\zeta^4+7\zeta^2+7^2} \left[-1\pm i\sqrt{3}\,\frac{\zeta^2+7}{\zeta^2-7}\right].
\label{eq_sup:A0Bpm}
%=\frac{6}{7}\,\frac{-1\pm i \sqrt{3}\coth\left(\phi\right)}{1+2\cosh\left(2\phi\right)}.
\end{eqnarray}
Using Eqs. (\ref{eq:unb_Four_k3})-(\ref{eq_sup:A0Bpm}) and noting that for $z\rightarrow 0$ we have $\zeta\simeq 7 (qz/6)^{1/3}/3 $, one finds that $\widetilde{\widehat{Q}}(\xi,0)=e^{i\xi n_0}$, and that $\widetilde{\widehat{Q}}(0,z)=A_0/(\lambda_0-1)+B_+/(\lambda_+-1)+B_-/(\lambda_--1)=(1-z)^{-1}$, indicating, respectively, that the occupation probability reduces to the initial condition at $t=0$ and that the propagator is normalised in space.

The propagator generating function in unbounded space, given by
\begin{align}
\widetilde{P}^{(3)}_{n_0}(n,z)=\frac{3}{8\pi zq}\int_{-\pi}^{\pi}\text{e}^{-i\xi(n-n_0)}\left\{\frac{A_0}{\lambda_0-\cos(\zeta)}+\frac{B_+}{\lambda_+-\cos(\zeta)}+\frac{B_-}{\lambda_--\cos(\zeta)} \right\}\text{d}\xi,
\label{eq_sup:Four_sol}
\end{align}
is used, analogously to the $k\in \{1,2\}$ cases, to derive via the method of images the bounded propagator
\begin{align}
&\widetilde{Q}^{(3)}_{n_0}(n,z)=\frac{1}{N(1-z)}+\frac{3}{4zqN}\left\{A_0\sum_{k=1}^{N-1}\frac{\cos\left[\frac{2\pi k(n-n_0)}{N}\right]}{\lambda_0-\cos\left(\frac{2\pi k}{N}\right)}+B_+\sum_{k=1}^{N-1}\frac{\cos\left[\frac{2\pi k(n-n_0)}{N}\right]}{\lambda_+-\cos\left(\frac{2\pi k}{N}\right)}+B_-\sum_{k=1}^{N-1}\frac{\cos\left[\frac{2\pi k(n-n_0)}{N}\right]}{\lambda_--\cos\left(\frac{2\pi k}{N}\right)}\right\} \nonumber \\
&= \frac{3}{4zq}\left\{A_0\frac{T_{N-|n-n_0|}(\lambda_0)+T_{|n-n_0|}(\lambda_0)}{\left(\lambda_0^2-1\right)U_{N-1}(\lambda_0)}+B_+\frac{T_{N-|n-n_0|}(\lambda_+)+T_{|n-n_0|}(\lambda_+)}{\left(\lambda_+^2-1\right)U_{N-1}(\lambda_+)}+B_-\frac{T_{N-|n-n_0|}(\lambda_-)+T_{|n-n_0|}(\lambda_-)}{\left(\lambda_-^2-1\right)U_{N-1}(\lambda_-)}\right\},
\label{eq_sup:finite_per_j31}
\end{align}
where, once again, identities linking finite trigonometric series to Chebyshev polynomials \cite{LucaPRX} have been used to obtain the last expression.

\section{Mean first-passage for the $K$-neighbour walk on the ring lattice}

We present here explicit expressions for the random walk  mean first-passage time (MFPT) on the ring lattice. For the cases presented above, that is for $k\in\{1,2,3\}$, we show how the MFPT
\begin{equation}
    \mathcal{F}^{(k)}_{n_0\to n} = \frac{k}{q}\sum_{\ell=1}^{N-1}\frac{\left[1-\cos\left(\frac{2\pi\ell(n-n_0)}{N}\right)\right]\sin\left(\frac{\ell \pi}{N}\right)}{ k \sin\left(\frac{\ell \pi}{N}\right)-\sin\left(\frac{k \ell \pi }{N}\right)\cos\left(\frac{(k+1)\ell \pi}{N}\right)},
    \label{eq_sup: MFPT_anyk}
\end{equation}
which we have slightly rewritten from Eq. (4) of the main text, i.e., here we consider the $q\in (0,1]$ case, can be expressed in an alternative form, that is we are able to perform the finite summation. We also consider the extreme case when $K$ covers the entire ring, that is when the walker may jump from one lattice site to any other lattice site.

\subsection{The nearest-neighbour ($k = 1$) walk}
In this case Eq. (\ref{eq_sup: MFPT_anyk}) reduces to
\begin{equation}
    \mathcal{F}^{(1)}_{n_0\to n} = \frac{1}{q}\sum_{\ell=1}^{N-1}\frac{1-\cos\left(\frac{2\pi\ell(n-n_0)}{N}\right)}{ 1-\cos\left(\frac{2\ell \pi}{N}\right)},
    \label{eq_sup: MFPT}
\end{equation}
which has been shown in ref. \cite{LucaPRX} to be equal to
%\begin{equation}
%   \sum_{\ell = 1}^{N-1}\frac{\cos\left(\frac{2\pi \ell m}{N}\right)}{1-\cos\left(\frac{2\pi k}{N}\right))} = \frac{1}{6}(N^2 - 1) + |m|(|m|-N) 
%\end{equation}
%and 
%\begin{equation}
%    \sum_{\ell = 1}^{N-1}\frac{1}{\sin^2\left(\frac{\pi \ell}{N}\right)} = \frac{N^2 - 1}{3},
%\end{equation}
%one can evaluate Eq. \ref{eq_sup: MFPT} exactly as 
\begin{equation}
    \mathcal{F}^{(1)}_{n_0\to n} = \frac{1}{q}\Big(N- |n-n_0|\Big)|n-n_0|.
\end{equation}
\subsection{The next-nearest-neighbour ($k = 2$) walk}
For this case, we exploit the explicit expression of the propagator derived in the last line of Eq. (\ref{eq_sup:sol_nnn_space_2}), rather than Eq. (2) of the main text, and via the renewal relation \cite{montroll1965random, redner2001guide}, $\widetilde{F}^{(2)}_{n_0}(n,z)=\frac{\widetilde{Q}^{(2)}_{n_0}(n,z)}{\widetilde{Q}^{(2)}_{n}(n,z)}$, we obtain the first-passage generating function
\begin{align}
&\widetilde{F}^{(2)}_{n_0}(n,z)=\Bigg\{\Big[T_{N-|n-n_0|}(\beta^{-}_z)+T_{|n-n_0|}(\beta^{-}_z)\Big]\Big[(\beta^+_z)^2-1\Big]U_{N-1}(\beta^+_z)\Bigg. \nonumber \\
&\Bigg.+(-1)^{n-n_0}\Big[T_{N-|n-n_0|}(\beta^{+}_z)+(-1)^NT_{|n-n_0|}(\beta^{+}_z)\Big]\Big[(\beta^-_z)^2-1\Big]U_{N-1}(\beta^-_z)\Bigg\} \nonumber \\
&\times \Bigg\{\Big[T_{N}(\beta^{-}_z)+1\Big]\Big[(\beta^+_z)^2-1\Big]U_{N-1}(\beta^+_z)+\Big[T_{N}(\beta^{+}_z)+(-1)^N\Big]\Big[(\beta^-_z)^2-1\Big]U_{N-1}(\beta^-_z)\Bigg\}^{-1}.
\label{eq:fppk2}
\end{align}

Via $\mathcal{F}^{(2)}_{n_0\rightarrow n}=\left.\frac{d}{dz}\widetilde{F}^{(2)}_{n_0}(n,z)\right|_{z=1}$ we obatin the MFPT
\begin{align}
&\mathcal{F}^{(2)}_{n_0\rightarrow n}=\frac{2}{5\,q}\Big(N-|n-n_0|\Big)|n-n_0|+\frac{4\sqrt{5}}{25\,q}N\left(\frac{\cosh\left[N\ln\left(\frac{3+\sqrt{5}}{2}\right)\right]+(-1)^N}{\sinh\left[N\ln\left(\frac{3+\sqrt{5}}{2}\right)\right]} \right. \nonumber \\
&\Bigg.\times\left\{1-(-1)^{n-n_0}\cosh\left[|n-n_0|\ln\left(\frac{3+\sqrt{5}}{2}\right)\right]\right\}+(-1)^{n-n_0}\sinh\left[|n-n_0|\ln\left(\frac{3+\sqrt{5}}{2}\right)\right]\Bigg).
\label{eq:mfptk2}
\end{align}
Given that the sum and subtraction of $\cosh$ and $\sinh$ terms grow very fast with $n$, it is numerically more stable to rewrite Eq. (\ref{eq:mfptk2}) as
\begin{align}
&\mathcal{F}^{(2)}_{n_0\rightarrow n}\!=\!\frac{2}{5\,q}\Big(N-|n-n_0|\Big)|n-n_0|\!+\!\frac{4\sqrt{5}}{25\,q}N\left(1-(-1)^{n-n_0}\left(\frac{2}{3+\sqrt{5}}\right)^{|n-n_0|}\right. \nonumber \\
&\left.+2\frac{(-1)^N+\left(\frac{2}{3+\sqrt{5}}\right)^N}{1-\left(\frac{2}{3+\sqrt{5}}\right)^N}\left\{\left(\frac{2}{3+\sqrt{5}}\right)^N\!-\!\frac{(-1)^{n-n_0}}{2}\left[\left(\frac{2}{3+\sqrt{5}}\right)^{N\!-\!|n-n_0|}\!+\!\left(\frac{2}{3+\sqrt{5}}\right)^{N\!+\!|n-n_0|}\right]\right\}\right).
\label{eq:mfptk3}
\end{align}

%As the term $\left[1-\frac{\cosh\left[N\ln\left(\frac{3+\sqrt{5}}{2}\right)\right]+(-1)^N}{\sinh\left[N\ln\left(\frac{3+\sqrt{5}}{2}\right)\right]}\right]\ll 1$ for $N\geq 8$, one can approximate Eq (\ref{eq:mfptk2}) extremely well with the simpler expression
%\begin{align}
%&\mathcal{F}^{(2)}_{n_0\rightarrow n}=\frac{2}{5\,q}\Big(N-|n-n_0|\Big)|n-n_0|+\frac{4\sqrt{5}}{25\,q}N\Bigg( 1-(-1)^{n-n_0}\left(\frac{2}{3+\sqrt{5}}\right)^{|n-n_0|}\Bigg),
%\label{eq:mfptk2_approx}
%\end{align}
%which shows clearly the departure from the $2(N-|n-n_0|)|n-n_0|/(5q)$ dependence.

\subsection{The next-next-nearest-neighbour ($k=3$) walk}
To find the MFPT from the limit $z\rightarrow 1$ of $\frac{d}{dz}\left[\widetilde{Q}^{(3)}_{n_0}(n,z)/\widetilde{Q}^{(3)}_{n}(n,z)\right]$ in this case, it is more convenient to assume that $z$ is a real variable and rewrite the coefficients $A_0$, $B_{\pm}$, $\lambda_0$, and $\lambda_{\pm}$. This can be done by defining 
\begin{align}
\alpha=1+\frac{3}{4q}\left[\frac{1}{z}-1\right],
\label{eq:alpha}
\end{align}
and finding the roots of the depressed cubic polynomial
\begin{align}
\alpha-\frac{5}{54}-y^3+\frac{7}{12}y=0,
\label{eq:alpha_eq}
\end{align}
obtained by substituting $y=\cos(\xi)+1/6$ in the denominator of the third line of Eq. (\ref{eq:unb_Four_k3}), that is in $\alpha-\cos^3(\xi)-\cos^2(\xi)/2+\cos(\xi)/2=0$. The roots of Eq. (\ref{eq:alpha_eq}) can be expressed with hyperbolic functions and provide the alternative form of the elements in Eq. (\ref{eq_sup:finite_per_j31}) when $z$ is real and positive:
\begin{align}
\lambda_0&=-\frac{1}{6}+\frac{\sqrt{7}}{3} \cosh\left(\upsilon\right), \\ \nonumber 
\lambda_{\pm}&=-\frac{1}{6}-\frac{\sqrt{7}}{6}  \left[\cosh\left(\upsilon\right)\mp i\sqrt{3} \sinh\left(\upsilon\right)\right],\\ \nonumber 
\upsilon&=\frac{1}{3}\mbox{arccosh}\left(\frac{108\,\alpha-10}{7^{\frac{3}{2}}}\right), \\ \nonumber
A_0&=\frac{12}{7}\frac{1}{1+2\cosh(2\upsilon)}, \nonumber \\
B_{\pm}&=-\frac{6}{7}\frac{1\mp i \sqrt{3}\coth(\upsilon)}{1+2\cosh(2\upsilon)}.
 \label{eq:sol_confined_b}
\end{align}

In the limit $z\rightarrow 1$, we have that $\upsilon\rightarrow \upsilon^*=3^{-1}\mbox{arccosh}[2\sqrt{7}]$, which in turns give $\lambda_0\rightarrow 1$ when $z\rightarrow 1$ since  $\cosh(\mbox{arccosh}[2\sqrt{7}]/3)=\sqrt{7}/2$.  We  thus conveniently write the first-passage probability as
\begin{align}
&\widetilde{F}_{n_0}(n,z)=\frac{h(n,n_0,\upsilon)+(\lambda_0-1)U_{N-1}(\lambda_0)g(n,n_0,\upsilon)}{h(n,n,\upsilon)+(\lambda_0-1)U_{N-1}(\lambda_0)g(n,n,\upsilon)},
\label{eq:fppk3}
\end{align}
where
\begin{align}
h(n,n_0,\upsilon)&=A_0\!\frac{T_{N-|n-n_0|}(\lambda_0)\!+\!T_{|n-n_0|}(\lambda_0)}{\left(\lambda_0+1\right)}, \nonumber \\
g(n,n_0,\upsilon)&=B_+\frac{T_{N-|n-n_0|}(\lambda_+)+T_{|n-n_0|}(\lambda_+)}{\left(\lambda_+^2-1\right)U_{N-1}(\lambda_+)}\!+\!B_-\!\frac{T_{N-|n-n_0|}(\lambda_-)\!+\!T_{|n-n_0|}(\lambda_-)}{\left(\lambda_-^2-1\right)U_{N-1}(\lambda_-)}.
\label{eq:gnn0z}
\end{align}

The terms that survive in the evaluation of the mean first-passage time $\mathcal{F}^{(3)}_{n_0 \to n}$ gives
\begin{align}
&\mathcal{F}^{(3)}_{n_0 \to n}=\left.\frac{d \upsilon }{dz}\right|_{z=1}\left\{\left[\frac{d  }{d\upsilon}h(n,n_0,\upsilon)\right]h(n,n,\upsilon)-\left[\frac{d }{d\upsilon}h(n,n,\upsilon)\right]h(n,n_0,\upsilon)\right. \nonumber \\
&\left.+\left(\frac{d}{d\upsilon}\Big[(\lambda_0-1)U_{N-1}(\lambda_0)\Big]\right)\Big[g(n,n_0,\upsilon)h(n,n,\upsilon)-g(n,n,\upsilon)h(n,n_0,\upsilon)\Big]\right\} \left.\Big\{h(n,n,\upsilon\Big\}^{-2}\right|_{\upsilon=\upsilon^*}
\label{eq:mfptk3_0}
\end{align}
Quantities relevant to evaluate the above MFPT are the following
\begin{align}
&\left.\frac{d \upsilon }{dz}\right|_{z=1}=-\frac{1}{q}\left(\frac{3}{7}\right)^{3/2} \nonumber \\
&h\left(n,n_0,\upsilon^*\right)=\frac{2}{7} \nonumber \\
&\left.\frac{d}{d \upsilon}h\left(n,n_0,\upsilon\right)\right|_{\upsilon=\upsilon^*}\!=\frac{1}{2\sqrt{21}}\big[5-N^2-2(N-|n-n_0|)|n-n_0|\big]
%\!\frac{\sqrt{3}\Big\{(7\!+\!\sqrt{7})\Big[N^2-2N|n-n_0|+2(n-n_0)^2\Big]\!-\!(19+\!\sqrt{7})\Big\}}{14(4+\sqrt{7})} 
\nonumber \\
&\left.\frac{d}{d\upsilon}\Big[(\lambda_0-1)U_{N-1}(\lambda_0)\right|_{\upsilon=\upsilon^*}=\frac{\sqrt{7}}{3}N\sinh\left(\upsilon^*\right)=\sqrt{\frac{7}{3}}\frac{N}{2} \nonumber \\
&g\left(n,n_0,\upsilon^*\right)=\frac{\sqrt{6}}{7}\frac{\sqrt{1+\frac{i}{3\sqrt{7}}}\Big\{\cosh[(N-|n-n_0|)\omega]+\cosh[(n-n_0)\omega]\Big\}}{\sinh\left(N\omega\right)}+\mbox{c.c.}
\label{eq:gnn0z1}
\end{align}
where $\omega=\mbox{arcosh}\left(-\frac{3}{4}-i\frac{\sqrt{7}}{4}\right)$.
This allows to write
\begin{equation}
\mathcal{F}^{(3)}_{n_0\rightarrow n}=\frac{3}{14q}(N-|n-n_0|)|n-n_0|+\frac{3N}{4q}\Big[g(n,n,\upsilon^*)-g(n,n_0,\upsilon^*)\Big].
\label{eq:MFPTkeq3}
\end{equation}

To extract directly the real part of Eq. (\ref{eq:MFPTkeq3}), one converts to polar coordinates in the complex domain and rewrite
\begin{align}
\sqrt{1\pm\frac{i}{3\sqrt{7}}}&=\frac{1}{\sqrt{2}(63)^{1/4}}\left[\sqrt{8+3\sqrt{7}}\pm i\sqrt{8-3\sqrt{7}}\right],\nonumber \\
\mbox{arcosh}\left(-\frac{3}{4}\pm i\frac{\sqrt{7}}{4}\right)&=\frac{1}{2}\ln\left(\frac{2+\sqrt{7}+\sqrt{7+4\sqrt{7}}}{2}\right)\pm i\left[\pi -\arctan\left(\frac{\sqrt{7+4\sqrt{7}}}{3}\right)\right]
\end{align}
and using the following definition
\begin{align}
f_1(m)=\cosh(m\rho)\cos(m\beta),\nonumber \\
f_2(m)=\cosh(m\rho)\sin(m\beta),\nonumber \\
f_3(m)=\sinh(m\rho)\cos(m\beta),\nonumber \\
f_4(m)=\sinh(m\rho)\sin(m\beta),
\label{eq_sup:funcT_def}
\end{align}
with $\rho=\frac{1}{2}\ln\left(\frac{2+\sqrt{7}+\sqrt{7+4\sqrt{7}}}{2}\right)$, $\beta=\arctan\left(\frac{\sqrt{7+4\sqrt{7}}}{3}\right)$,
and $m\geq 0$ an integer. The ensuing identity
\begin{align}
&\left(\sqrt{8+3\sqrt{7}}+i\sqrt{8-3\sqrt{7}}\right)\cosh\left\{m\,\mbox{arccosh}\left[\frac{-3+i\sqrt{7}}{4}\right]\right\}\sinh\left\{N\,\mbox{arccosh}\left[\frac{-3-i\sqrt{7}}{3}\right]\right\}\nonumber \\
&+\left(\sqrt{8+3\sqrt{7}}-i\sqrt{8-3\sqrt{7}}\right)\cosh\left\{m\,\mbox{arccosh}\left[\frac{-3-i\sqrt{7}}{4}\right]\right\}\sinh\left\{N\,\mbox{arccosh}\left[\frac{-3+i\sqrt{7}}{3}\right]\right\}\nonumber \\
&=2(-1)^{N+m}\left\{\sqrt{8\!+\!3\sqrt{7}}\bigg[f_1(m)f_3(N)+f_2(N)f_4(m)\bigg]\!-\!\sqrt{8\!-\!3\sqrt{7}}\bigg[f_1(m)f_2(N)-f_3(N)f_4(m)\bigg]\right\},
\label{eq:identity}
\end{align}
is employed to obtain finally
%\begin{align}
%&\mathcal{F}^{(3)}_{n_0\rightarrow n}=\frac{1}{q}\frac{3}{14}\Big(N-|n-n_0|\Big)|n-n_0| \nonumber \\
%&+\frac{N}{q}\frac{3}{7^{5/4}}\frac{\mathcal{G}(N)+(-1)^N\mathcal{G}(0)-(-1)^{n-n_0}\Big[\mathcal{G}(N-|n-n_0|)+(-1)^N\mathcal{G}(|n-n_0)|)\Big]}{\cosh\left[N\ln\left(\frac{2+\sqrt{7}+\sqrt{7+4\sqrt{7}}}{2}\right)\right]-\cos\left[N\ln\left(\frac{2+\sqrt{7}+\sqrt{7+4\sqrt{7}}}{2}\right)\right]},
%\label{eq:MFPTkeq3_old}
%\end{align}
\begin{align}
&\mathcal{F}^{(3)}_{n_0\rightarrow n}=\frac{1}{q}\frac{3}{14}\Big(N-|n-n_0|\Big)|n-n_0| \nonumber \\
&+\frac{N}{2q}\frac{3}{7^{5/4}}\frac{\mathcal{G}(N)+(-1)^N\mathcal{G}(0)-(-1)^{|n-n_0|}\Big[\mathcal{G}(N-|n-n_0|)+(-1)^N\mathcal{G}(|n-n_0)|)\Big]}{1+\sigma^{4N}-2\sigma^{2N}\cos\left[2N\ln(\sigma)\right]},
\label{eq:MFPTkeq3_b}
\end{align}
where
\begin{align}
%&\mathcal{G}(m)=4\sqrt{8\!+\!3\sqrt{7}}\bigg[f_1(m)f_3(N)+f_2(N)f_4(m)\bigg]\!-\!4\sqrt{8\!-\!3\sqrt{7}}\bigg[f_1(m)f_2(N)-f_3(N)f_4(m)\bigg]\nonumber \\
&\mathcal{G}(m)=\sigma^{N-m}\sqrt{8+3\sqrt{7}}\Big\{\Big(\sigma^{2m}-\sigma^{2N}\Big)\cos\Big[(N+m)\beta\Big]+\Big(1-\sigma^{2(m+N)}\Big)\cos\Big[(N-m)\beta\Big]\Big\}\nonumber \\
&-\sigma^{N-m}\sqrt{8-3\sqrt{7}}\Big\{\left(1+\sigma^{2(N+m)}\right)\sin\Big[(N+m)\beta\Big]+\Big(\sigma^{2N}+\sigma^{2m}\Big)\sin\Big[(N-m)\beta\Big]\Big\}
\label{eq:mathcalg}
\end{align}
with $\sigma=\sqrt{\frac{2}{2+\sqrt{7}+\sqrt{7+4\sqrt{7}}}}.$

\subsection{The lattice walk MFPT on the all-to-all graph for odd $N$}

A simple way to derive the expression for this case from Eq. (\ref{eq_sup: MFPT_anyk}) is to consider $N$ odd and to take $k = (N-1)/2$. One then obtains
\begin{equation}
    \mathcal{F}_{n_0 \to n}^{\left(\frac{N-1}{2}\right)} = \frac{1}{q}\sum_{\ell=1}^{N-1}\frac{\left[1-\cos\left(\frac{2\pi\ell(n-n_0)}{N}\right)\right]}{1-\frac{2}{N-1}\frac{\sin\left(\frac{(N-1) \ell \pi }{2N}\right)\cos\left(\frac{(N+1)\ell \pi}{2N}\right)}{\sin\left(\frac{\ell \pi}{N}\right)}}.
    \label{eq_sup: MFPTathcal{F}_limk}
\end{equation}
Exploiting the relation $\sin(A)\cos(B) = \frac{\sin(A+B)+\sin(A-B)}{2}$ it is straightforward to simplify Eq. (\ref{eq_sup: MFPTathcal{F}_limk}) to 
\begin{equation}
    \mathcal{F}^{\left(\frac{N-1}{2}\right)}_{n_0\to n} = \frac{N-1}{Nq}\sum_{\ell=1}^{N-1}\left[1-\cos\left(\frac{2\pi\ell(n-n_0)}{N}\right)\right],
    \label{eq_sup: MFPTathcal{F}_limk1}
\end{equation}
where the finite summation is now trivial to evaluate leading to 
\begin{equation}
    \mathcal{F}^{\left(\frac{N-1}{2}\right)}_{n_0\to n} = \frac{1}{q}(1-\delta_{n,n_0})(N-1)
\end{equation}
given in the main text. To consider also the case when $N$ is even, we have first derived the dynamics of the occupation probability for any $N$ and then extracted the MFPT as we show below.

\section{Occupation probability dynamics for the lattice walk on the all-to-all graph for odd and even $N$}

We start by identifying the lattice walk transition matrix $\mathbb{S}$ for the all-to-all network. Its elements are $\mathbb{S}_{i,i}=1-q$, which represents the probability to stay at a node, while all the off-diagonal elements, which describe the probability to move to any of the other nodes are $\mathbb{S}_{i,j}=\frac{q}{N-1}$ for $i\not=j$, dictated by conservation of probability. Such a matrix can be conveniently expressed as
\begin{align}
\mathbb{S}=\frac{q}{N-1}\mathbb{T}
\label{eq_sup:trans_mat}
\end{align}
%\begin{align}
%&\mathbb{S}=\left(\begin{array}{ccccc} 1-q & \frac{q}{N-1} & ... & \frac{q}{N-1} & \frac{q}{N-1} \\
% \frac{q}{N-1} & 1-q & ... & \frac{q}{N-1} & \frac{q}{N-1} \\
%... & ... & ... & ... & ... \\
% \frac{q}{N-1} & \frac{q}{N-1} & ... & 1-q &\frac{q}{N-1}     \\
% \frac{q}{N-1} & \frac{q}{N-1} & ... & \frac{q}{N-1}   & 1-q 
%\end{array}\right)=\frac{q}{N-1}\mathbb{T}
%\label{eq_sup:trans_mat}
%\end{align}
with $\mathbb{T}_{i,i}=(N-1)(1-q)/q$ and $\mathbb{T}_{i,j}=1$ for $i\not= j$. 
%\begin{align}
%\mathbb{T}=\left(\begin{array}{ccccc} \frac{(N-1)(1-q)}{q} & 1 & ... & 1 & 1\\
% 1 & \frac{(N-1)(1-q)}{q}  & ... & 1 & 1 \\
%... & ... & ... & ... & ... \\
%1 & 1 & ... & \frac{(N-1)(1-q)}{q}  &1     \\
% 1 & 1 & ... & 1  & \frac{(N-1)(1-q)}{q} 
%\end{array}\right)
%\end{align}
The Master equation in vectorial form is given by
\begin{equation}
\boldsymbol{\phi}(t+1)=\mathbb{S}\cdot \boldsymbol{\phi}(t),
\label{eq_sup: all-to_allME}
\end{equation}
where we use $\phi(n,t)$ to denote the lattice walk occupation probability when constrained to the all-to-all graph. Equation \ref{eq_sup: all-to_allME} is solved through eigendecomposition as
\begin{align}
\boldsymbol{\phi}(t)=\mathbb{S}^t\cdot \boldsymbol{\phi}(0)=\left(\frac{q}{N-1}\right)^t\mathbb{T}^t\cdot \boldsymbol{\phi}(0)=\left(\frac{q}{N-1}\right)^t\left(\boldsymbol{A}\cdot\boldsymbol{\Lambda}\cdot\boldsymbol{A}^{-1}\right)^t\cdot \boldsymbol{\phi}(0)=\left(\frac{q}{N-1}\right)^t\boldsymbol{A}\cdot\boldsymbol{\Lambda}^t\cdot\boldsymbol{A}^{-1}\cdot \boldsymbol{\phi}(0),
\label{eq_sup:gen_sol}
\end{align}
where $\boldsymbol{\Lambda}$ is the diagonal matrix of eigenvalues of $\mathbb{T}$, namely $\boldsymbol{\Lambda}_{ij}=0$ for $i\not= j$, $\boldsymbol{\Lambda}_{11}=(N-1)/q$ and $\boldsymbol{\Lambda}_{ii}=[N(1-q)-1]/q$ for $i=2,...,N$. $\boldsymbol{A}$ is the corresponding matrix of normalised right eigenvectors $\boldsymbol{A}$, whose elements are
 $\boldsymbol{A}_{i,1}=-\frac{1}{\sqrt{N}}$ for $i=2,...,N$,  $\boldsymbol{A}_{1,j}=\frac{1}{\sqrt{N}}$ for $j=1,...,N$, $\boldsymbol{A}_{i+1,N+1-i}=\frac{1}{\sqrt{N}}$ for $i=1,...,N-1$ and 0 elsewhere. For the inverse we have instead $\boldsymbol{A}^{-1}_{1,j}=\frac{1}{\sqrt{N}}$, $\boldsymbol{A}^{-1}_{i+1,N+1-i}=\frac{N-1}{\sqrt{N}}$ for $i=1,...,N-1$ and $-\frac{1}{\sqrt{N}}$ elsewhere.

%\begin{tiny}
%\begin{equation}
%\boldsymbol{\Lambda}=\left(\begin{array}{ccccccc} \frac{(N-1)(1-q)-1}{q} & 0 & 0 & ... & 0 & 0& 0 \\
% 0 & \frac{(N-1)(1-q)-1}{q}  & 0 & ... & 0 & 0 & 0 \\
%0 & 0 & \frac{(N-1)(1-q)-1}{q}  &... & 0 & 0 & 0 \\
%... & ... & ... & ... &  ... & ... & ... \\
%0 & 0& 0 & ... & \frac{(N-1)(1-q)-1}{q}  & 0 &0 \\
%0 & 0 & 0 & ... &0 &\frac{(N-1)(1-q)-1}{q}  &0    \\
%0 & 0 & 0 & ...   & 0 &0 &\frac{N-1}{q} 
%\end{array}\right),
%\end{equation}
%\end{tiny}
%\begin{equation}
%\boldsymbol{A}=\frac{1}{\sqrt{N}}\left(\begin{array}{ccccccc} -1 & 0 & 0 & ... & 0 & 0& 1 \\
% -1 & 0 & 0 & ... & 0 & 1 & 0 \\
%-1 & 0 & 0 &... & 1 & 0 & 0 \\
%... & ... & ... & ... &  ... & ... & ... \\
%-1 & 0& 1 & ... & 0 & 0 &0 \\
%-1 & 1 & 0 & ... &0 &0 &0    \\
%1 & 1 & 1 & ...   & 1 &1 &1
%\end{array}\right),
%\end{equation}
%and its inverse
%\begin{equation}
%\boldsymbol{A}^{-1}=\frac{1}{\sqrt{N}}\left(\begin{array}{ccccccc} -1 & -1 & -1 & ... & -1 & -1 & N-1 \\
% -1 & -1 & -1 & ... & -1 & N-1 & 1 \\
%-1 & -1 & -1 &... & N-1 & -1 & 1 \\
%... & ... & ... & ... &  ... & ... & ... \\
%-1 & -1& N-1 & ... & -1 & -1 &1 \\
%-1 & N-1 & -1 & ... &-1 &-1 &1    \\
%N-1 & -1 & -1 & ...   & -1 &-1 &1
%\end{array}\right).
%\end{equation}
After a bit of algebra and defining $\boldsymbol{A}^{-1}\cdot\boldsymbol{\Lambda}^t\cdot\boldsymbol{A}=\mathbb{K}$ one gets
\begin{align}
\mathbb{K}_{i,i}=\frac{1}{N}\left[\left(\frac{N-1}{q}\right)^t+(N-1)\left(\frac{(N-1)(1-q)}{q}-1\right)^t\right],
\label{eq_sup:matK_jk}
\end{align}
and for $i\not=j$
\begin{align}
\mathbb{K}_{i,j}=\frac{1}{N}\left[\left(\frac{N-1}{q}\right)^t-\left(\frac{(N-1)(1-q)}{q}-1\right)^t\right].
\label{eq_sup:matK_ji}
\end{align}
Inserting (\ref{eq_sup:matK_jk}) and (\ref{eq_sup:matK_ji}) in (\ref{eq_sup:gen_sol}) and considering a starting condition with a walker localised at $n=n_0$, that is $\phi(n,0)=\delta_{n,n_0}$, one obtains the simple result
\begin{equation}
\phi_{n_0}(n,t)=\frac{1}{N}\left[1-\left(1-q-\frac{q}{N-1}\right)^t\right]+\left(1-q-\frac{q}{N-1}\right)^t\delta_{n,n_0},
\label{eq_sup:sol_time}
\end{equation}
which is properly normalised over all the nodes of the network and satisfy the initial condition. Note that for the case $q=(N-1)/N$, which corresponds to the jump probabilities being all equal including the self loops, Eq. (\ref{eq_sup:sol_time}) reduces to the trivial dynamics
\begin{equation}
\phi_{n_0}(n,t)=\frac{1}{N}\left[1-\delta_{t,0}\right]+\delta_{t,0}\delta_{n,n_0},
\label{eq_sup:sol_time_2}
\end{equation}

\subsection{First-passage probability and MFPT}

It is a trivial exercise now to take the generating function of (\ref{eq_sup:sol_time}), namely
\begin{align}
\widetilde{\phi}_{n_0}(n,z)&=\frac{1}{N}\left[\frac{1}{1-z}-\frac{1}{1-z\left(1-q-\frac{q}{N-1}\right)}\right]+\frac{1}{1-z\left(1-q-\frac{q}{N-1}\right)}\delta_{n,n_0} \nonumber \\
&=\frac{1}{N}\left[\frac{1}{1-z}+\frac{N\delta_{n,n_0}-1}{1-z\left(1-q-\frac{q}{N-1}\right)}\right]
\label{eq_sup:sol_time_z}
\end{align}
and write for $n\not= n_0$ the first-passage probability generating function as
 \begin{equation}
\widetilde{F}^{(\mbox{\tiny{all}})}_{n_0}(n,z)=\frac{\widetilde{\phi}_{n_0}(n,z)}{\widetilde{\phi}_{n}(n,z)}=\frac{q}{N-1}\frac{z}{1-z\left(1-\frac{q}{N-1}\right)}.
\label{eq_sup:f_p_z}
\end{equation}
Inverting Eq. (\ref{eq_sup:f_p_z}) to time gives
\begin{align}
F^{(\mbox{\tiny{all}})}_{n_0}(n, t)=\frac{q}{N-1}\left(1-\frac{q}{N-1}\right)^{t-1}, \qquad t\ge 1,
\label{eq_sup:fp_time}
\end{align}
and $F^{(\mbox{\tiny{all}})}_{n_0}(n, 0)=0$, which, as expected, is independent of $n$. For the special case where all jump probabilities are equal, i.e. $q=(N-1)/N$, Eq. (\ref{eq_sup:fp_time}) reduces to
\begin{equation}
   F^{(\mbox{\tiny{all}})}_{n_0}(n, t)=\frac{\left(1-1/N\right)^{t-1}}{N}, \qquad t\ge 1,
\end{equation}
while the case $q=1$ gives
\begin{equation}
    F^{(\mbox{\tiny{all}})}_{n_0}(n, t)=\frac{1}{N-1}\left(\frac{N-2}{N-1}\right)^{t-1}, \qquad t\ge 1,
\end{equation}
which was reported (without derivation) in ref. \cite{sood2004first} as $\frac{1}{N-1}\left(\frac{N-2}{N-1}\right)^{t}$.

For the MFPT it is straightforward to obtain from (\ref{eq_sup:fp_time}) 
\begin{align}
\mathcal{F}^{(\mbox{\tiny{all}})}_{n_0\rightarrow n}=\frac{N-1}{q}.
\label{eq:mfptall}
\end{align}

\section{Occupation probability on the small-world network}
\label{sec:op_small_world}
In Eq. (6) of the main text we have presented the Master equation governing the dynamics of $S(n,t)$ on SWN, where the network is constructed by rewiring links of $K$-neighbour ring lattice. The rewiring makes the ring lattice spatially disordered, with spatial heterogeneities of the inert type (probability preserving). The rewiring are parameterised via a set of four-tuples $(u, v, \eta_{v, u}, \eta_{u,v})$ with $u$ and $v$ denoting a pair of sites where the outgoing connections between them are modified by $\eta_{v, u}$, $\eta_{u,v}$, respectively. With defective sites being of the inert type, the general formalism of ref. \cite{sarvaharman2023particle} that allows to describe the dynamics in disordered lattices can be employed. Through that formalism we derive analytically the  propagator of the Master equation as the generating function 
\begin{equation}
\begin{aligned}
    \widetilde{S}_{n_0}(n, z) &= \widetilde{Q}_{n_0}^{(k)}(n, z) - 1 + \frac{\det[{\mathbb{H}(n, n_0, z)}]}{\det[\mathbb{H}(z)]},
    \label{eq_sup: hetero_occu}
\end{aligned}
\end{equation}
where the matrices of size $M\times M$ are as follows
\begin{equation}
    \mathbb{H}(z)_{i, j}  =  \eta_{u_i, v_i}\widetilde{Q}_{\langle u_j -  v_j\rangle}^{(k)}(u_i, z) - \eta_{v_i, u_i}\widetilde{Q}_{\langle u_j -  v_j\rangle}^{(k)}(v_i, z) - \frac{\delta_{i, j}}{z},
    \label{eq_sup: kgen}
\end{equation}
and 
\begin{equation}
    \mathbb{H}(n, n_0, z)_{i, j} =  \mathbb{H}(z)_{i, j} - \widetilde{Q}_{\langle u_j -  v_j\rangle}^{(k)}(n, z)\left[\eta_{u_i, v_i}\widetilde{Q}_{n_0}^{(k)}(u^{\phantom{\prime}}_i, z) - \eta_{v_i, u_i}\widetilde{Q}_{n_0}^{(k)}(v_i, z)\right].
    \label{eq_sup: kn0gen}
\end{equation}
Note that the difference of the (defect-free) occupation probability evaluated at sites where the transition probabilities have been modified enter the evaluation of $\widetilde{S}_{n_0}(n, z)$, and it is apparent in terms represented through the notation $f_{\langle u - v\rangle}(\cdot) = f_{u}(\cdot) - f_{v}(\cdot)$.

\begin{figure}[h!]
    \includegraphics[width = \textwidth]{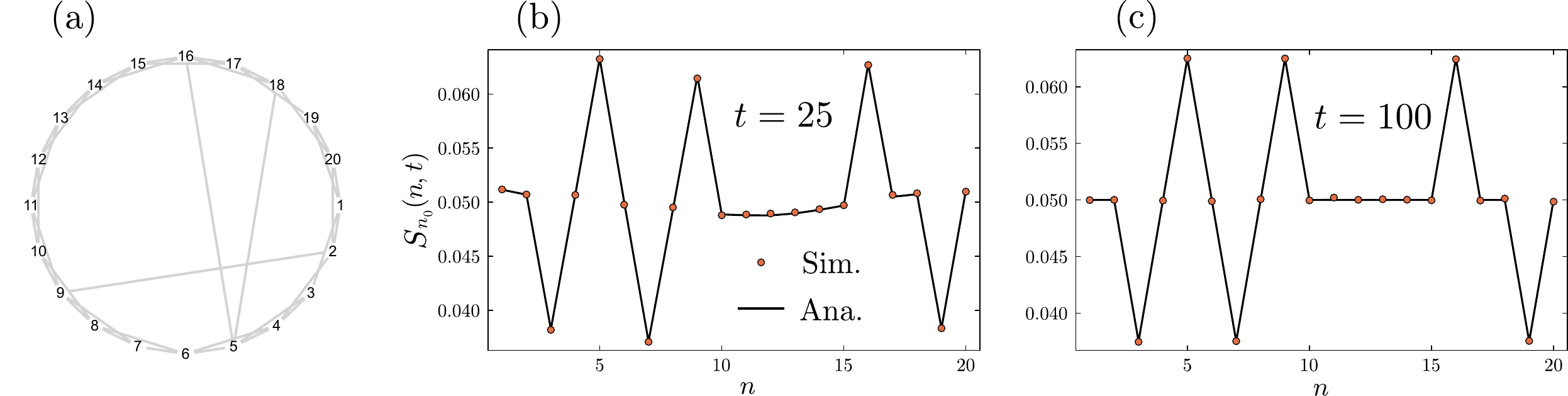}
    \caption[A comparison between the theoretically predicted occupation probability and a simulation at two timesteps
    in a small-world network with 20 nodes.]{A comparison between the theoretically predicted occupation probability, extracted from Eq.
    \ref{eq_sup: hetero_occu}, and the average of $10^6$ simulations at (b) $t=25$ and (c) $t=100$ in a specific SWN realisation generated from a $K=4$ ring lattice with $p=0.15$ and $N=20$ depicted in (a). The initial condition is $n_0 = 1$.}
    \label{fig: occu_sw}
\end{figure}
In Fig. \ref{fig: occu_sw} we compare against stochastic simulations the dynamics of $S_{n_0}(n,t)$ extracted by inverting numerically in time through a straightforward one-dimensional trapezoidal integration \cite{abate1992fourier} of Eq. (\ref{eq_sup: hetero_occu}). The plot shows that the occupation probability in the SWN
is qualitatively very different from the broadening of a single peaked function centred on $n_0$ that one would
expect at short times on the ring lattice. $S_{n_0}(n,t)$ instead displays a complex dependence on the degree
of the node and the initial condition, with some nodes with probability values already approaching the steady state after only 25 steps, while some nodes that are near or in correspondence of the nodes with long range connections remain farther from it. Note that for the steady state we have \cite{noh2004random} $S_{n_0}(n,t\rightarrow \infty)=\chi_n/\mathcal{E}$, with $\chi_n$ the coordination number of node $n$ and $\mathcal{E}$ the number of edges in the network.

\section{Mean first-passage and first-absorption time on the small-world network}
The MFPT on the network can also be derived algebraically from the occupation probability in Eq. (\ref{eq_sup: hetero_occu}).   The implementation of the general MFPT formalism from ref. \cite{sarvaharman2023particle}) in the SWN context gives
\begin{equation}
	\mathdutchcal{F}^{(k)}_{n_0\to n} = \frac{\mathcal{F}^{(k)}_{n_0\to n}\det\left[\mathbb{L} - \left(1/\mathcal{F}^{(k)}_{n_0\to n}\right)\mathbb{M}\right]}{\det\left[\mathbb{L}-\mathbb{O}\right]},  
    \label{eq_sup: hetero_MFPT}
\end{equation}
where the matrices of size $M\times M$ are
\begin{align}
\mathbb{L}_{i,j} &= \frac{\eta_{v_i, u_i}}{\mathcal{R}^{(k)}_{u_i}} \mathcal{F}^{(k)}_{\langle u_j - v_j \rangle \to u_i}
- \frac{\eta_{u_i, v_i}}{\mathcal{R}^{(k)}_{v_i}} \mathcal{F}^{(k)}_{\langle u_j - v_j \rangle \to v_i}
+ \delta_{i,j},  \\
\mathbb{M}_{i,j} &= \left( 
\frac{\eta_{v_i, u_i}}{\mathcal{R}^{(k)}_{u_i}} \mathcal{F}^{(k)}_{\langle n_0 - n \rangle \to u_i}
- \frac{\eta_{u_i, v_i}}{\mathcal{R}^{(k)}_{v_i}} \mathcal{F}^{(k)}_{\langle n_0 - n \rangle \to v_i}
\right)
\mathcal{F}^{(k)}_{\langle u_j - v_j \rangle \to n},  \\
\mathbb{O}_{i,j} &= \left( 
\frac{\eta_{v_i, u_i}}{\mathcal{R}^{(k)}_{u_i}} - \frac{\eta_{u_i, v_i}}{\mathcal{R}^{(k)}_{v_i}} 
\right) \mathcal{F}^{(k)}_{\langle u_j - v_j \rangle \to n}. 
\end{align}
Note that the MFPT $\mathdutchcal{F}^{(k)}_{n_0\to n}$ depends on the disorder parameters and the MFPT, but also on the mean return time (MRT), in the ring lattice.

When the target is partially absorbing ($0<\rho<1$), the above MFPT expression is used to calculate the mean-first absorption time (MFAT) from $n_0$ to $n$, whose general expression is given by (see e.g. \cite{giuggiolieta2013,giuggioli2022spatio})
\begin{align}
\mathcal{A}_{n_0\to n}(\rho) = \mathdutchcal{F}^{(k)}_{n_0\to n} + \frac{1-\rho}{\rho} \mathcal{R}_{n},
\label{eq:MFAT}
\end{align}
with the MRT being $\mathcal{R}_n=\mathcal{E}/\chi_{n}$.  

\section{First-passage probability and mean first-absorption time for the defective ring lattice}

As described in the main text, we consider a ring lattice with each lattice site having $K$ links to the nearest neighbours and may include one self loop into itself. We modify such lattice to create a special network by placing one short-cut between nodes $n+1$ and $n_0+5$, which requires the use of $M= K+1$ defects in the the exact formalism described in Sec. \ref{sec:op_small_world}. Using Eq. (\ref{eq_sup: hetero_occu}) for $  \widetilde{S}_{n_0}(n, z)$ we are able to study the dynamics of first-absorption from site $n_0$ to site $n$ in this special network via
\begin{align}
\widetilde{A}_{n_0}(n, z) = \frac{\rho\,   \widetilde{S}_{n_0}(n, z)}{1-\rho + \rho\,  \widetilde{S}_{n}(n, z)}.
\label{eq:first_abs_gen}
\end{align}
\begin{figure}[hb]
    \centering
    \includegraphics[width=0.5\linewidth]{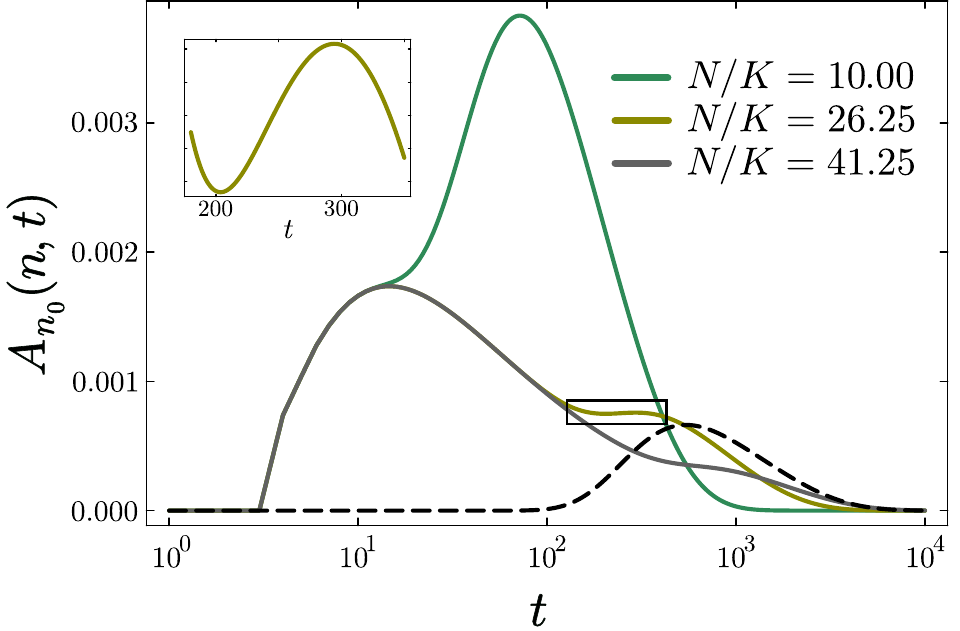}
    \caption{Temporal dependence of the first-absorption probability from $n_0=1$ to the partially absorbing site ($\rho=0.85$) at $n=N/2$ for the above described network with $K=8$ and different values of $N$. The dashed curve represents the case without any rewiring of the original ring lattice. The inset is a blow up of the region $180<t<340$ for the $N/K=26.25$  case.} 
    \label{fig:time_fa}
\end{figure}
We numerical invert to time  \cite{abate1992fourier} Eq. (\ref{eq:first_abs_gen}) and plot the first-absorption dynamics for different choice of $N$ and $\rho$ in Fig. \ref{fig:time_fa}. For one parameter selection, namely $N/K=10.00$, we have chosen $N$ small enough so that it falls just below the regime of bi-modality (see Fig. 3(b) in the main text). One may notice that a change of flex after $t=10$ steps, which is indicative of the fact the network is so small that the trajectories that eventually get absorbed at the target site and take the shortcut and those that move around the ring are of comparable timescales, which gives rise to the pronounced mode. For a bigger network, namely $N=210$ ($N/K=26.25$), we enter the bi-modal regime, and the above timescales can now be differentiated: the network is large enough for the first mode to appear in correspondence of the change of flex of the network with $N=80$. The third choice of network size, namely $N=330$ ($N/K=41.25$) is above the bi-modal region and one observe that the mode due to the direct trajetcories is still present, but the second one has disappeared. (For other choices of $N$ and $\rho$ we have been presented plots in Fig. 3(c) of the main text.)

\begin{figure}[ht]
    \centering
    \includegraphics[width=0.5\linewidth]{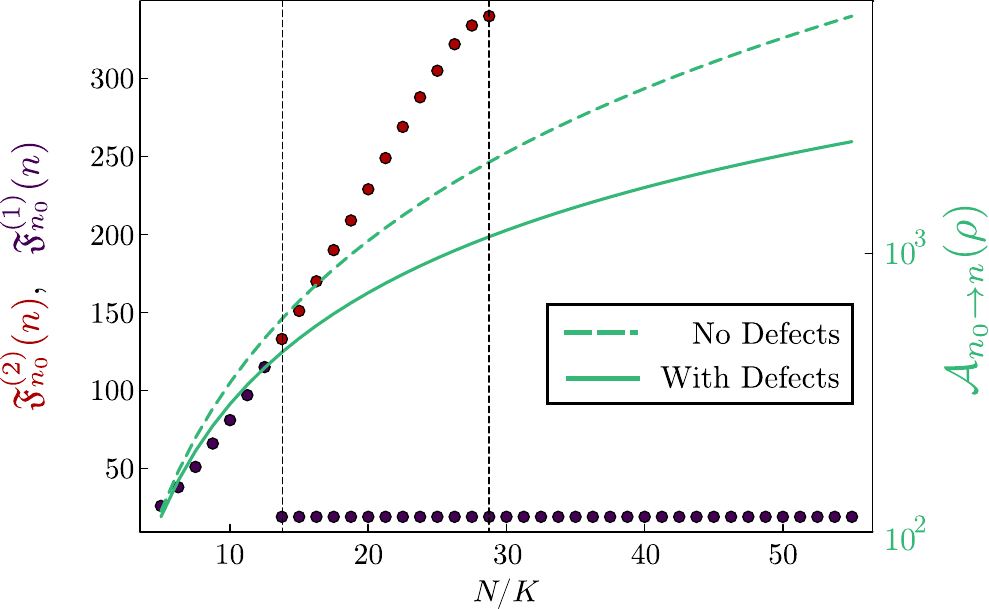}
    \caption{Dependence of the first and second mode of the first-absorption probability, respectively, $\mathfrak{F}^{(1)}_{n_0}(n)$ and $\mathfrak{F}^{(2)}_{n_0}(n)$, and the MFAT in Eq. (\ref{eq:MFAT}) for the small world network selected in the main text in Fig. 3 when $\rho = 0.5$ and $K=8$ as a function of the network size $N$. The choice of $n_0$ and $n$ are described in Fig. \ref{fig:time_fa}. The specific choice of the network and the target location $n$ gives $\mathcal{R}_n=N.$ The region in between the two vertical dashed lines corresponds to when both modes are present. The green solid represents the MFAT in the network in the absence of the short-cut, that is the MFAT in the ring lattice.} 
    \label{fig:mfat_versus_N}
\end{figure}

The analytic formula in Eq (\ref{eq:MFAT}) allows us to study the mean first-absorption time as a function of the network size and compare it to the value of the first and second mode of the first-absorption probability. While in the main text we have shown such comparison in Fig 3(d) when $\rho=1$, in Fig. \ref{fig:mfat_versus_N} we present the case $\rho=0.5$. A comparison to the corresponding plot in the main text, shows a rightward shift and a narrowing of the region of bimodality, which is consistent with what observed in Fig. 3(b) of the main text.

\end{document}